\documentclass[aps,pre,twocolumn,superscriptaddress,floatfix]{revtex4-1}

\usepackage{graphicx}
\usepackage[latin1]{inputenc}
\usepackage{amsfonts}
\usepackage{amssymb}
\usepackage{amsmath,amssymb}
\usepackage{bm}
\usepackage{cancel}
\usepackage{cleveref}
\usepackage{subfigure} 
\usepackage{bbold}
\usepackage{caption}
\usepackage{color}

\begin{document}
\graphicspath{{images/}}

\title{Active Ornstein-Uhlenbeck particles}
\author{L.L. Bonilla}
\affiliation{G. Mill\'an Institute for Fluid Dynamics, Nanoscience \& Industrial Mathematics, and Department of Materials Science \& Engineering and Chemical Engineering, Universidad Carlos III de Madrid, Avenida de la Universidad 30, 28911 Legan\'es, Spain. Email: {\tt bonilla@ing.uc3m.es}}
\affiliation{Courant Institute of Mathematical Sciences, New York University, 251 Mercer St, New York, N.Y. 10012, USA}
\date{\today}

\begin{abstract}
Active Ornstein-Uhlenbeck particles (AOUPs) are overdamped particles in an interaction potential subject to external Ornstein-Uhlenbeck noises. They can be transformed into a system of underdamped particles under additional velocity dependent forces and subject to white noise forces. There has been some discussion in the literature on whether AOUPs can be in equilibrium for particular interaction potentials and how far from equilibrium they are in the limit of small persistence time. By using a theorem on the time reversed form of the AOUP Langevin-Ito equations, I prove that they have an equilibrium probability density invariant under time reversal if and only if their smooth interaction potential has zero third derivatives. In the limit of small persistence Ornstein-Uhlenbeck time $\tau$, a Chapman-Enskog expansion of the Fokker-Planck equation shows that the probability density has a local equilibrium solution in the particle momenta modulated by a reduced probability density that varies slowly with the position. The reduced probability density satisfies a continuity equation in which the probability current has an asymptotic expansion in powers of $\tau$. Keeping up to $O(\tau)$ terms, this equation is a diffusion equation, which has an equilibrium stationary solution with zero current. However, $O(\tau^2)$ terms contain fifth and sixth order spatial derivatives and the continuity equation no longer has a zero current stationary solution. The expansion of the overall stationary solution now contains odd terms in the momenta, which clearly shows that it is not an equilibrium.
\end{abstract}

\maketitle

\section{Introduction}\label{sec:1}
In a recent paper, Fodor {\em et al} pose the question of how far from equilibrium is active matter by considering overdamped active Ornstein-Uhlenbeck particles (AOUPs) \cite{fodor}. AOUPs subject to a short range repulsive potential exhibit a clustering phase transition when their density is sufficiently large compared to noise features. Since the Ornstein-Uhlenbeck noise becomes white noise as its correlation (persistence) time $\tau$ decreases, it is legitimate to investigate whether overdamped AOUPs are near equilibrium for small persistence times. Fodor {\em et al} argue that, for small $\tau$, AOUPs are in an extended equilibrium state characterized by a modified energy and by a nonzero production of entropy that is proportional to $\tau^2$ times a third derivative of the potential energy \cite{fodor}. AOUPs in a harmonic potential are thus in equilibrium \cite{fodor}. The production of entropy in the stationary state follows from a formula that involves the entropy of direct and time reversed AOUP paths and an ergodicity assumption \cite{gas04,and07,spi12}. For small persistence times and arbitrary smooth potential, there is a nonthermal AOUP equilibrium that breaks down at higher order in $\tau$ \cite{fodor}.  

This picture has been disputed by Mandal {\em et al} \cite{mandal} who calculated the AOUP production of entropy and found it to be nonzero even for quadratic potential energy \cite{mandal}. Their production of entropy is based on a formula for the time reversed stochastic process of the AOUP (which is derived in the supplementary material of Ref.~\cite{mandal}) and on formulas for the energy and heat transfer. This result was, in turn, declared to be incorrect by Caprini {\em et al}, who used a different calculation of entropy production also based on path integral representations \cite{cap_comment,pug17}. Further discussions of entropy calculations, including the convenience to change the model by adding a thermal noise to the AOUPs, are found in Ref.~\cite{dab19} and references therein.

In this paper, I prove that AOUPs in a smooth quadratic potential indeed reach thermal equilibrium at an effective temperature and calculate the corresponding probability density. AOUPs in potentials with nonzero third derivatives reach a nonequilibrium stationary state whose probability density is not invariant under time reversal. The proof by contradiction is based on a theorem that yields the drift and diffusion of the time reversed process for a given Langevin-Ito equation \cite{hau86}. The importance of this theorem for stochastic control and its roots in earlier results of Edward Nelson's (cf. Chapter 13 in Ref.~\cite{nel67}) have been recently emphasized by Chen {\em et al} \cite{che15}. Remarkably, the time reversed stochastic diffusion process derived by Mandal {\em et al} \cite{mandal} does not satisfy the theorem in Ref.~\cite{hau86}. 

I also consider how AOUPs are approximately in an equilibrium state in the limit of small persistence time, $\epsilon \propto \sqrt{\tau}\to 0$. I derive an equation for the reduced probability density limit of small persistence time by using the Chapman-Enskog method \cite{BT10}. To leading and first order, including $O(\epsilon^2)$ terms, it is a diffusive Smoluchowski equation \cite{gardiner}, although consideration of the Chapman-Enskog method as an expansion in gradients \cite{cha70} would suggest that third and fourth order derivatives should have also appeared. This second order diffusive Smoluchowski equation has a stationary equilibrium solution whose probability current vanishes. The next order terms in the Chapman-Enskog expansion of the equation for the reduced probability density are $O(\epsilon^4)$ and contain fifth and sixth order spatial derivatives, which suggest the stationary solution to be out of equilibrium. In fact, to the same order of approximation, the stationary solution of the corresponding Fokker-Planck equation (FPE) is no longer invariant under time reversal, which confirms it as a nonequilibrium state.

The rest of the paper is as follows. In section \ref{sec:2}, I recall the AOUP model, adapt the time reversal theorem to it, and prove the main result. For completeness, the time reversal theorem is enunciated in Appendix \ref{ap1}, cf. Ref.~\cite{hau86} for technical details and proof. For the sake of simplicity and to minimize obfuscation, I consider in section \ref{sec:2} a single particle and give the corresponding formulas for the general case in Appendix \ref{ap2}. In section \ref{sec:3}, I derive equations for the reduced probability density of a single AOUP in the limit of small persistence time. In this limit, momenta equilibrate rapidly whereas coordinates evolve in a slower scale, just as in the Smoluchowski approximation to the FPE for overdamped particles. I show that the stationary probability current vanishes up to leading and first order in the persistence time. The extension of these results to a system of many AOUPs is by no means obvious. Details for the general case of $N$ AOUPs are given in Appendices \ref{ap3} (reduced equation) and \ref{ap4} (approximate equilibrium probability density). The case of a single AOUP in a harmonic potential under the action of an additional white noise source is considered in Appendix \ref{ap5}. Section \ref{sec:4} contains the conclusions of this work. 

\section{Equilibrium for an active Ornstein-Uhlenbeck particle}\label{sec:2}
I consider a single AOUP particle in this section and give the details about a system of $N$ AOUPs in Appendix \ref{ap2}. The equation of motion for one AOUP is \cite{fodor}
\begin{equation}
\dot{x}=-\mu\Phi'(x)+v,\quad \tau\dot{v}=-v+\sqrt{2D}\eta(t), \label{eq1}
\end{equation}
where $\Phi(x)$ is the potential energy and $\mu$ the mobility of the particle. $\eta(t)$ is a zero-mean delta-correlated white noise. From the definition $p=\dot{x}$, it follows
\begin{equation*}
\dot{p}=\dot{v}-\mu\Phi''(x)\dot{x}=-\frac{v}{\tau}-\mu\Phi''(x)p+\sqrt{\frac{2D}{\tau^2}}\eta(t).
\end{equation*}
Thus, I have obtained the system of equations
\begin{eqnarray}
&&\dot{x}=p,    \label{eq2}\\
&&\dot{p}=-\frac{p+\mu\Phi'(x)}{\tau}-\mu\Phi''(x)p+\sqrt{\frac{2D}{\tau^2}}\eta(t).\label{eq3}
\end{eqnarray}
The corresponding FPE for the probability density $\rho(x,p,t)$ is 
\begin{equation}
\frac{\partial \rho}{\partial t}\!+\!p\frac{\partial\rho}{\partial x} \!-\! \frac{\partial}{\partial p}\!\left[\!\left(\frac{1\!+\!\mu\tau\Phi''}{\tau}\, p \!+\!\frac{\mu}{\tau} \Phi' \right)\! \rho \!+\!\frac{D}{\tau^2}\frac{\partial \rho}{\partial p}\right]\!=\!0. \label{eq4}
\end{equation}

In Ref.~\cite{mandal}, it is argued that the time reversed process of the solution of Eqs.~\eqref{eq2}-\eqref{eq3} satisfies 
\begin{eqnarray}
\dot{x}=p,    \quad \dot{p}=-\frac{p+\mu\Phi'(x)}{\tau}+\mu\Phi''(x)p+\sqrt{\frac{2D}{\tau^2}}\eta(t). \quad \label{eq5}
\end{eqnarray}
What does this mean? The stochastic process $(x,p)$ starts from some initial condition $(x_0,p_0)$ and, at time $t_f$, stops at some random value $(x_f,p_f)$. The time reversed process should start at time $t'=t_f-t=0$ at $(x_f,p_f)$ and then go back randomly to exactly $(x_0,p_0)$ at $t'=t_f$ as the solution of some stochastic differential equation. It seems astonishing that this time reversal is achieved by Eq.~\eqref{eq5}, which does not use  information from the direct process given by Eqs.~\eqref{eq2}-\eqref{eq3}. In fact, the astonishing Eq.~\eqref{eq5} is not true. 

According to Eqs.~\eqref{a2}-\eqref{a4} in Appendix \ref{ap1}, the reverse time stochastic differential equations of Eqs.~\eqref{eq2}-\eqref{eq3} for $dt>0$ are \cite{nel67,hau86,che15}
\begin{eqnarray}
&&\dot{\overline{x}}=\overline{p},   \label{eq6}\\
&&\dot{\overline{p}}=-\frac{\overline{p}[1+\mu\tau\Phi''(\overline{x})]}{\tau}-\frac{\mu}{\tau} \Phi'(\overline{x})-\frac{2D}{\tau^2}\frac{\partial\ln\rho}{\partial p}(\overline{x},\overline{p},t)\nonumber\\
&&\quad+\sqrt{\frac{2D}{\tau^2}}\overline{\eta}(t),\label{eq7}
\end{eqnarray}
where $\rho$ is the solution of the FPE \eqref{eq4}, with initial condition $\rho(x_0,p_0,0)$, and $\overline{\eta}(t)$ is the derivative of a standard Wiener process $\hat{w}(t)$ whose past $\{\hat{w}(s); 0\leq s\leq t\}$ is independent of $(\overline{x},\overline{p})$ for all $t\geq 0$ \cite{che15}, cf. Appendix \ref{ap1}. Note that the drift term in the time reversed equation, Eq.~\eqref{eq7}, depends on the time dependent solution of the forward time FPE \eqref{eq4} $\rho(\overline{x},\overline{p},t)$, with $\overline{x}$, $\overline{p}$ taking values on the time reversed process.

Let me assume now that the probability density of Eqs.~\eqref{eq2}-\eqref{eq3} evolves to an equilibrium state whose density, $\rho_s(x,p)$, is invariant under time reversal. This means that the drift term in Eq.~\eqref{eq7} for $\rho=\rho_s$ has to be the same as the drift term in Eq.~\eqref{eq3} under the time reversal transformation that leaves Eq.~\eqref{eq2} invariant: $t\to-t$, $\overline{x}=x$, $\overline{p}=-p$,
\begin{eqnarray*}
-\frac{p(1+\mu\tau\Phi'')+\mu\Phi'}{\tau}-\frac{2D}{\tau^2}\frac{\partial\ln\rho_s}{\partial p}=\frac{p(1+\mu\tau\Phi'')-\mu\Phi'}{\tau}.
\end{eqnarray*}
This yields
\begin{eqnarray}
\frac{2D}{\tau^2}\frac{\partial\ln\rho_s}{\partial p}=-\frac{2}{\tau} p (1+\mu\tau\Phi'')\Longrightarrow \nonumber\\
\rho_s=\exp\!\left[\Lambda(x)-\frac{\tau p^2}{2D}(1+\mu\tau\Phi'')\right]\!.\label{eq8}
\end{eqnarray}
I now insert this stationary probability density in the FPE \eqref{eq4} thereby finding
\begin{eqnarray*}
\rho_s p\left(\Lambda' +\frac{\mu}{D}\Phi'+\frac{\mu^2\tau}{D}\Phi'\Phi''\right)\! -\frac{\mu\tau^2\rho_s}{2D}p^3\Phi''' =0, 
\end{eqnarray*}
where the first term can be cancelled by choosing $\Lambda=-\frac{\mu}{D}\Phi-\frac{\mu^2\tau}{2D} \Phi'^2$ but not the second (unless $\Phi'''=0$). Thus there is an equilibrium state
\begin{eqnarray}
\rho_s=\frac{1}{Z}\exp\!\left[-\frac{\mu}{D}\!\left(\Phi +\frac{\mu\tau}{2}\Phi'^2\right)\!-\frac{\tau p^2}{2D}(1+\mu\tau\Phi'')\right]\!,\label{eq9}
\end{eqnarray}
with $\int \rho_sdx\, dp=1$, if and only if $\Phi''' =0$, which occurs for a smooth quadratic potential. I conclude that the stationary solution of the FPE is not an equilibrium, invariant under time reversal, unless $\Phi$ is quadratic, e.g., $\Phi=\kappa x^2/2$. In this case, the probability density is Gaussian:
\begin{eqnarray}
\rho_s(x,p) = \frac{1}{Z}\exp\!\left[-\frac{1}{T}\left(\frac{\kappa x^2}{2}+\frac{\tau p^2}{2\mu}\right)\!\right]\!,  \label{eq10}\\
 T=\frac{D}{\mu(1+\mu\tau\kappa)},\quad Z=2\pi T\sqrt{\frac{\mu}{\kappa\tau}}.  \nonumber
\end{eqnarray}
Eq.~\eqref{eq10} is the equilibrium probability density for a particle of mass $\mu/\tau$ placed in a harmonic potential and in contact with a bath at temperature $T=D/[\mu(1+\mu\tau\kappa)]$. Undoing the transformation, $p=v-\mu\kappa x$, it is immediate to prove that Eq.~\eqref{eq10} is also the equilibrium solution of the FPE for $\rho$ in the variables $x$ and $v$. As shown in Appendix \ref{ap2}, this result also holds for a system of $N$ AOUPs.

In Ref.~\cite{fodor}, Fodor {\em et al} also found that AOUPs in a quadratic potential reach thermal equilibrium at temperature $T=D/\mu$ (if $\tau=0$). By using path integrals, they showed that the production of entropy in the stationary state vanishes for a smooth quadratic $\Phi$. For such a potential, the stationary probability is Gaussian, as indicated by Eq.~\eqref{eq10}. Surprisingly in view of Eq.~\eqref{eq10}, Mandal {\em et al} have claimed that there is a positive production of entropy for AOUPs with any potential, including smooth quadratic potentials \cite{mandal}. However, their arguments are based upon incorrectly deducing that the time reversed process of Eqs.~\eqref{eq2}-\eqref{eq3} are Eq.~\eqref{eq5}, i.e., Eqs.~(7a)-(7b) in Ref.~\cite{mandal}. See the supplementary material in Ref.~\cite{mandal}, where Mandal {\em et al} implement their time reversal procedure. 

\section{Derivation of a reduced equation for small persistence time}\label{sec:3}
In this section, I derive a continuity equation for the reduced probability density of a single AOUP in the limit of small persistence time $\tau\propto\epsilon^2$. It is relatively simple to obtain a leading order approximation but I will derive an approximate equation including $O(\epsilon^5)$ terms. Keeping up to $O(\epsilon^3)$ terms in the continuity equation, there is an equilibrium solution with zero probability current. No equilibrium solution exists beyond this order, as I show by direct calculation of the approximate stationary state.

Firstly, let me nondimensionalize the FPE~\eqref{eq4} according to Table \ref{table1}. 
\begin{table}[ht]
\begin{center}\begin{tabular}{cccccc}
 \hline
$x$& $p$ & $t$ &$\Phi$& $\rho$\\ 
$l$ & $\sqrt{\frac{D}{\tau}}$ & $l\sqrt{\frac{\tau}{D}}$ &$\frac{D}{\mu}=T$ & $\sqrt{\frac{\tau}{Dl^2}}$\\ 
\hline
\end{tabular}
\end{center}
\caption{Units for nondimensionalizing the AOUP FPE \eqref{eq4}. $l$ is a characteristic length.}
\label{table1}
\end{table}
The nondimensional FPE is  
\begin{eqnarray}
\frac{\partial}{\partial p}\!\left(p\rho\!+\!\frac{\partial\rho}{\partial p}\right)\! =\! \epsilon\!\left[\frac{\partial \rho}{\partial t}\!+\!p\frac{\partial\rho}{\partial x} \!-\! \Phi'\frac{\partial\rho}{\partial p} \!-\!\epsilon \Phi'' \frac{\partial(p\rho)}{\partial p}\right]\!,\quad \label{eq11}
\end{eqnarray}
where the diffusive length is much smaller than the characteristic particle length $l$:
\begin{equation}
\epsilon=\frac{\sqrt{D\tau}}{l}\ll 1. \label{eq12}
\end{equation}
By an abuse of notation, I have kept the same symbols for dimensional and nondimensional variables. Table \ref{table1} can be used to get dimensional variables from the corresponding nondimensional ones. For $N$ particles in a $d$-dimensional cubic box of size $L$, one can use $l=(L^d/N)^{1/d}$, cf. Appendix \ref{ap3}. In this section, $N=d=1$. The limit $\epsilon\to 0$ corresponds to $\sqrt{\tau}\to 0$ in Fodor {\em et al}'s paper \cite{fodor}.  For $\epsilon=0$, Eq.~\eqref{eq11} has the solution
\begin{eqnarray}
\rho^{(0)}(x,p,t)= \frac{e^{-p^2/2}}{\sqrt{2\pi}}R(x,t;\epsilon), \label{eq13}\\
 \int \rho^{(0)}(x,p,t)\, dp= R(x,t;\epsilon).\nonumber
\end{eqnarray}

\subsection{Chapman-Enskog derivation of the reduced equation}
Given the stated goal of obtaining a high order approximation of the reduced equation for $R$, it is convenient to use the {\em Chapman-Enskog method} \cite{BT10}. I consider 
\begin{eqnarray}
&& \rho(x,p,t;\epsilon)=\frac{e^{-p^2/2}}{\sqrt{2\pi}}R(x,t;\epsilon)+\sum_{j=1}^\infty\epsilon^j \rho^{(j)}(x,p;R),\quad \label{eq14}\\
&&\frac{\partial R}{\partial t}= \sum_{j=1}^\infty\epsilon^j\mathcal{F}^{(j)}(R),\label{eq15}
\end{eqnarray}
where, for $j\geq 1$, 
\begin{equation}
\int \rho^{(j)}(x,p;R)\, dp=0. \label{eq16}
\end{equation}
The key ingredient of the Chapman-Enskog method is that the equation for $R$ in Eq.~\eqref{eq15} is expanded, not its solution. The functionals $\mathcal{F}^{(j)}(R)$ are calculated by imposing that the resulting linear  equations for the $\rho^{(j)}$ have solutions. If one keeps more than one term in Eq.~\eqref{eq15}, then this reduced equation for $R$ contains higher order terms that can regularize its leading order. For the original application to derive hydrodynamics from the Boltzmann equation, the leading order equations are the Euler equations and the equations including first order terms are the Navier-Stokes equations \cite{cha70}. For applications to unfolding  degenerate bifurcations in different contexts, including synchronization of oscillators and active matter, see Refs.~\cite{BT10,bon00,BT19}. 

I now proceed with the mechanics of the Chapman-Enskog method. The normalization condition for the probability density yields
\begin{equation}
\int R(x,t;\epsilon)\, dx=1. \label{eq17}
\end{equation}
Note that integrating Eq. \eqref{eq11} over the momenta, I obtain the exact continuity equation \begin{equation}
\frac{\partial R}{\partial t}+ \frac{\partial J}{\partial x}= 0, \quad J(x,t;\epsilon)=\int p\, \rho(x,p,t;\epsilon)\, dp.   \label{eq18}
\end{equation}
Inserting \eqref{eq13}-\eqref{eq15} into \eqref{eq11}, I obtain the hierarchy of equations
\begin{eqnarray}
&&\mathcal{L}\rho^{(1)}\equiv \frac{\partial}{\partial p}\!\left(p+\frac{\partial}{\partial p}\right)\!\rho^{(1)}= \frac{e^{-p^2/2}}{\sqrt{2\pi}}\, p\!\left(\Phi'+\frac{\partial}{\partial x} \right)\!R,\quad \label{eq19} \\
&& \mathcal{L}\rho^{(2)}=\frac{e^{-p^2/2}}{\sqrt{2\pi}}[\mathcal{F}^{(1)}+ (p^2-1)\Phi'' R] +p \frac{\partial\rho^{(1)}}{\partial x}\nonumber\\
&& \quad -\Phi'\frac{\partial\rho^{(1)}}{\partial p}, \label{eq20}\\
&& \mathcal{L}\rho^{(3)}\!=
p\frac{\partial\rho^{(2)}}{\partial x}\! -\!\Phi' \frac{\partial\rho^{(2)}}{\partial p}\!-\!\Phi'' \frac{\partial(p\rho^{(1)})}{\partial p}
\!+\!\frac{\delta\rho^{(1)}}{\delta R}\mathcal{F}^{(1)}\!, \label{eq21}\\
&& \mathcal{L}\rho^{(4)}=\frac{e^{-p^2/2}}{\sqrt{2\pi}}\mathcal{F}^{(3)} +p\frac{\partial\rho^{(3)}}{\partial x}-\Phi' \frac{\partial\rho^{(3)}}{\partial p}\nonumber\\
&& \quad-\Phi''\frac{\partial(p\rho^{(2)})}{\partial p} +\frac{\delta\rho^{(2)}}{\delta R}\mathcal{F}^{(1)}, \label{eq22}\\
&& \mathcal{L}\rho^{(5)}=
p\frac{\partial\rho^{(4)}}{\partial x} -\Phi' \frac{\partial\rho^{(4)}}{\partial p}-\Phi'' \frac{\partial(p\rho^{(3)})}{\partial p} +\frac{\delta\rho^{(1)}}{\delta R}\mathcal{F}^{(3)}\nonumber\\
&& \quad+\frac{\delta\rho^{(3)}}{\delta R}\mathcal{F}^{(1)}, \label{eq23}
\end{eqnarray}
etc. The solvability conditions for the equations of this hierarchy are that the integrals over momenta of their right hand sides be zero. I have used that $\mathcal{F}^{(2j)}=0$ (see below) and kept enough equations in the hierarchy to obtain the reduced equation including terms beyond those corresponding to an equilibrium state.

\subsubsection{Result: Diffusion equation to order $\epsilon^3$}
As I show below, up to $O(\epsilon^5)$ terms, $R$ is the solution of the Smoluchowski diffusion equation:
\begin{eqnarray}
\frac{\partial R}{\partial t}=\epsilon\frac{\partial}{\partial x}\!\left(R\Phi'+\frac{\partial}{\partial x}[(1-\epsilon^2\Phi'')R]\right)\!+ O(\epsilon^5).\label{eq24}
\end{eqnarray}
One would have expected this equation to contain terms with three and four derivatives with respect to $x$ because, after all, the Chapman-Enskog expansion is an ``expansion in gradients'' \cite{cha70}. Such terms cancel out for the AOUP in Eq.~\eqref{eq24}, which then has the extended equilibrium solution $e^{-\tilde{\Phi}}/Z$, with $\tilde{\Phi}=\Phi+\epsilon^2(\Phi'^2/2-\Phi'')$. See also Ref.~\cite{fodor}. 

\subsubsection{Derivation of the probability current}
The solutions $\rho^{(j)}$ of equations in the hierarchy \eqref{eq19}-\eqref{eq23} are Gaussians in $p$ times polynomials of degree $j$: 
\begin{eqnarray}
\rho^{(2j+\xi)}(x,p;R)=\frac{e^{-p^2/2}}{\sqrt{2\pi}}\sum_{n=0}^j\mathcal{A}^{(2j+\xi)}_{2n+\xi}p^{2n+\xi}, \label{eq25}
\end{eqnarray}
where $\xi=0,1$. Clearly for \eqref{eq20}-\eqref{eq23}, the solvability conditions yield
\begin{equation}
\mathcal{F}^{(j)}=-\frac{\partial J^{(j)}}{\partial x},\quad J^{(j)} =\int p\rho^{(j)} dp.  \label{eq26}
\end{equation} 
Eq.~\eqref{eq25} then implies that $J^{(2j)}=0$ and $\mathcal{F}^{(2j)}=0$, which I have used to suppress all such terms in the previous hierarchy of equations. Note that, in order to find terms of order $\epsilon^5$ in Eq.~\eqref{eq24}, I need to solve Eqs.~\eqref{eq19}-\eqref{eq23}, but not higher order equations in the hierarchy. 

Eqs.~\eqref{eq15} and \eqref{eq26} agree with the continuity equation \eqref{eq18}. Using 
\begin{equation}
\mathcal{L}\!\left(\frac{p\, e^{-p^2/2}}{\sqrt{2\pi}}\right)\!=-\frac{p\, e^{-p^2/2}}{\sqrt{2\pi}},\label{eq27}
\end{equation}
The solution of Eq.~\eqref{eq19} that satisfies Eq.~\eqref{eq16} is:
\begin{equation}
\rho^{(1)}=- \frac{e^{-p^2/2}}{\sqrt{2\pi}} p \! \left(\Phi'R+\frac{\partial R}{\partial x} \right)\!\equiv - \frac{p\, e^{-p^2/2}}{\sqrt{2\pi}}\,\mathcal{D}R. \label{eq28}
\end{equation}
Inserting this in Eq.~\eqref{eq20}, its solvability condition produces
\begin{eqnarray}
&&\mathcal{F}^{(1)}=-\frac{\partial J^{(1)}}{\partial x},\nonumber\\
&& J^{(1)} =\int p\rho^{(1)}dp=-\mathcal{D}R=-e^{-\Phi}\frac{\partial}{\partial x}(e^\Phi R). \label{eq29}
\end{eqnarray}

Using
\begin{eqnarray}
&&\mathcal{L}\!\left(\frac{p^2 e^{-p^2/2}}{\sqrt{2\pi}} \right)\!= -\frac{e^{-p^2/2}}{\sqrt{2\pi}} 2(p^2-1), \label{eq30}\\
&&\int \frac{e^{-p^2/2}}{\sqrt{2\pi}}p^2 dp=1.\nonumber
\end{eqnarray}
I find the solution of Eq.~\eqref{eq20} that satisfies Eq.~\eqref{eq16}:
\begin{eqnarray}
&&\rho^{(2)}=\frac{e^{-p^2/2}}{\sqrt{2\pi}} (p^2-1)\, \mathcal{A}^{(2)},\label{eq31}\\
&&\mathcal{A}^{(2)}=\frac{1}{2}\!\left(\frac{\partial^2R}{\partial x^2}+2\Phi'\frac{\partial R}{\partial x}+ \Phi'^2 R\right)\!. \label{eq32}
\end{eqnarray}
Eq. \eqref{eq31} has the form of Eq.~\eqref{eq25} with $\mathcal{A}^{(2)}_0=-\mathcal{A}^{(2)}_2=-\mathcal{A}^{(2)}$. 
Using 
\begin{equation}
\mathcal{L}\!\left(\frac{p^3e^{-p^2/2}}{\sqrt{2\pi}}\right)\!=\frac{e^{-p^2/2}}{\sqrt{2\pi}}3p(2-p^2),\label{eq33}
\end{equation}
I find the solution of Eq.~\eqref{eq21} that satisfies Eq.~\eqref{eq16}: 
\begin{eqnarray}
&&\rho^{(3)}\!\!=\frac{p\, e^{-p^2/2}}{\sqrt{2\pi}}(\mathcal{A}^{(3)}_3 p^2 + \mathcal{A}^{(3)}_1) \nonumber\\
&&\quad=\! \frac{p\, e^{-p^2/2}}{\sqrt{2\pi}}[\mathcal{A}^{(3)}_3(p^2-3) \!+\!J^{(3)}] ,\label{eq34}\\
&&\mathcal{A}^{(3)}_3=-\frac{1}{3}(\mathcal{D}\mathcal{A}^{(2)} -\Phi'' \mathcal{D}R),\label{eq35}\\
&&\mathcal{A}^{(3)}_1 \! =\!\left(\!\mathcal{D} \frac{\partial}{\partial x}-2\Phi''\right)\!\mathcal{D}R +6\mathcal{A}^{(3)}_3 \!+ \!(2\Phi' +\mathcal{D})\mathcal{A}^{(2)}.\quad \label{eq36}
\end{eqnarray}
The corresponding contribution to the probability current is
\begin{eqnarray}
&&J^{(3)}=3\mathcal{A}^{(3)}_3+\mathcal{A}^{(3)}_1= \!\left(\mathcal{D}\frac{\partial}{\partial x} +\Phi'' \right)\!\mathcal{D}R - 2\frac{\partial\mathcal{A}^{(2)}}{\partial x}\nonumber\\
&&\Longrightarrow J^{(3)}\!=\frac{\partial}{\partial x} (\Phi''R). \label{eq37}
\end{eqnarray}

To find $\rho^{(4)}$, I follow the same procedure. The solution of Eq.~\eqref{eq22} with Eqs.~\eqref{eq31} and \eqref{eq34} is Eq.~\eqref{eq25} with
\begin{eqnarray}
&&\rho^{(4)}\!=\!\frac{e^{-p^2/2}}{\sqrt{2\pi}}[\mathcal{A}^{(4)}_4 (p^4-3)+\mathcal{A}^{(4)}_2 (p^2-1)], \label{eq38}\\
&&\mathcal{A}^{(4)}_4=-\frac{1}{4}(\mathcal{D}\mathcal{A}^{(3)}_3+ \Phi''\mathcal{A}^{(2)}),\label{eq39}\\
&&\mathcal{A}^{(4)}_2=\!\frac{1}{4}\!\left(\frac{\partial^2}{\partial x^2} +2\Phi'\frac{\partial}{\partial x}+\Phi'^2\right)\!\frac{\partial}{\partial x}\mathcal{D}R+\!\frac{3}{2}\Phi' \mathcal{A}^{(3)}_3
\nonumber\\
&&\quad\quad+\frac{1}{2}\Phi''\mathcal{A}^{(2)} -\frac{1}{2}\mathcal{D} J^{(3)},   \label{eq40}\\
&&\mathcal{A}^{(4)}_0= -3\mathcal{A}^{(4)}_4 -\mathcal{A}^{(4)}_2, \label{eq41}
\end{eqnarray}
where I have used
\begin{eqnarray}
&&\mathcal{L}\!\left(\frac{p^4e^{-p^2/2}}{\sqrt{2\pi}}\right)\!=\!\frac{4p^2e^{-p^2/2}}{\sqrt{2\pi}}(3-p^2),\nonumber\\
&&\quad\int \frac{e^{-p^2/2}}{\sqrt{2\pi}}p^4 dp=3.\label{eq42}
\end{eqnarray}

To find $\rho^{(5)}$, I insert Eqs.~\eqref{eq28}, \eqref{eq34} and \eqref{eq38} into Eq.~\eqref{eq23} and use the following formulas 
\begin{eqnarray}
&&\mathcal{L}\!\left(\frac{p^5e^{-p^2/2}}{\sqrt{2\pi}}\right)\!=\!\frac{5p^3e^{-p^2/2}}{\sqrt{2\pi}}(4-p^2), \nonumber\\
&&\quad \int \frac{p^5e^{-p^2/2}}{\sqrt{2\pi}}dp=15. \label{eq43}
\end{eqnarray}
to solve the resulting equation. The result is Eq.~\eqref{eq25} with 
\begin{eqnarray}
\mathcal{A}^{(5)}_5\!&=&\!-\frac{1}{5}(\mathcal{D}\mathcal{A}^{(4)}_4+ \Phi'' \mathcal{A}^{(3)}_3),\label{eq44}\\
\mathcal{A}^{(5)}_3\!&=&\! -\frac{4}{3}\frac{\partial}{\partial x}\mathcal{A}^{(4)}_4-\frac{1}{3}\mathcal{D}\mathcal{A}^{(4)}_2+\Phi''\!\left(\mathcal{A}^{(3)}_3-\frac{1}{3}J^{(3)}\right)\!\nonumber\\
&-&\frac{1}{3}\frac{\delta\mathcal{A}^{(3)}_3}{\delta R}\mathcal{D}R,  \label{eq45}\\
\mathcal{A}^{(5)}_1\!&=&\!\left(3\Phi'-5\frac{\partial}{\partial x}\right)\!\mathcal{A}^{(4)}_4 +\!\left(\Phi'-\frac{\partial}{\partial x} \right)\!\mathcal{A}^{(4)}_2\nonumber\\
&-&\mathcal{D}\frac{\partial J^{(3)}}{\partial x}+\frac{\delta\mathcal{A}^{(3)}_3}{\delta R} \frac{\partial}{\partial x}\mathcal{D}R -\frac{\delta J^{(3)}}{\delta R} \frac{\partial}{\partial x}\mathcal{D}R.\quad \label{eq46}
\end{eqnarray}

The probability current of Eq.~\eqref{eq26} that corresponds to Eqs.~\eqref{eq44}-\eqref{eq46} is
\begin{eqnarray}
J^{(5)}&=&3\frac{\partial^2}{\partial x^2} \mathcal{A}^{(3)}_3 +3\frac{\partial}{\partial x} (\Phi'' \mathcal{A}^{(2)})\nonumber\\
&-&\!\left(\frac{\delta\mathcal{A}^{(2)}}{\delta R}+\frac{\delta J^{(3)}}{\delta R}\right)\! \frac{\partial}{\partial x}\mathcal{D}R. \label{eq47}
\end{eqnarray}

The resulting reduced equation for the probability density $\rho$ is
\begin{eqnarray}
\frac{\partial R}{\partial(\epsilon t)}&=&\frac{\partial}{\partial x}\!\left[R\Phi'+\frac{\partial R}{\partial x} -\epsilon^2\frac{\partial}{\partial x}(R\Phi'')-\epsilon^4 J^{(5)}\right]\!\nonumber\\
&+& O(\epsilon^6)\equiv -\frac{\partial J^r}{\partial x}. \label{eq48}
\end{eqnarray}
Note that the reduced probability density evolves in a slow time scale $\epsilon t$. Eq.~\eqref{eq48} is a diffusion equation to $O(\epsilon^2)$ but $J^{(5)}$ contains derivatives of orders 3 to 5. 

\subsection{Equilibrium solution to $O(\epsilon^2)$}
There is an equilibrium solution that solves the reduced equation \eqref{eq48} with $J^r=0$ to order $\epsilon^5$ in the Chapman-Enskog expansion. To find it, I insert the exponential form
\begin{equation}
\rho_{\rm eq}=e^{\tilde{f}_{\rm eq}-\tilde{\Phi}}, \quad\tilde{\Phi}(x;\epsilon)=\Phi(x)+\sum_{j=1}^2 \epsilon^{2j}\Phi^{(2j)}(x),\label{eq49}
\end{equation}
into $J^r=0$ and find the $\Phi^{(2j)}$. The free energy $\tilde{f}_{\rm eq}$ ensures that the normalization condition \eqref{eq17} is fulfilled. To leading order, $J^{(1)}=0$ produces $\tilde{\Phi}=\Phi$ according to Eq.~\eqref{eq29}. Keeping the $\epsilon^2$ term in Eq.~\eqref{eq48}, I get
\begin{eqnarray*}
-\epsilon^2\!\left(\frac{\partial\Phi^{(2)}}{\partial x}+\Phi''' -\Phi'\Phi''\right)\!=O(\epsilon^4),
\end{eqnarray*}
which yields
\begin{equation}
R_{\rm eq}=e^{\tilde{f}_{\rm eq}-\tilde{\Phi}}, \quad\tilde{\Phi}=\Phi +\epsilon^2\!\left(\frac{1}{2}\!\Phi'^2-\Phi''\right)\!+O(\epsilon^4).\label{eq50}
\end{equation}
While $\Phi$ may be purely repulsive, the extra term in $\tilde{\Phi}$ of Eq.~\eqref{eq50} may produce an attractive component that is responsible for the segregation phase transition observed in Ref.~\cite{fodor}. The equilibrium probability density of Eq.~\eqref{eq50} can be generalized to the case of $N$ AOUPs as shown in Appendix \ref{ap3}. It coincides with Eq.~(7) of Ref.~\cite{fodor}.

For the term of order $\epsilon^4$, I get
\begin{eqnarray*}
\frac{\partial\Phi^{(4)}}{\partial x}&=& \frac{1}{2}\!\left(5\Phi''\Phi'''-\Phi'(\Phi'')^2 -\Phi'^2\Phi'''+ 2 \Phi' \frac{\partial^4\Phi}{\partial x^4}\right.\\
&-&\left.\frac{\partial^5\Phi}{\partial x^5}\right)\!. 
\end{eqnarray*}
Integrating this equation, I get
\begin{eqnarray}
\Phi^{(4)}&=& \frac{1}{2}\!\left(2\frac{\partial}{\partial x}(\Phi'\Phi'')-\Phi'^2\Phi'' -\frac{1}{2}\frac{\partial}{\partial x}(\Phi'')^2-\frac{\partial^4\Phi}{\partial x^4}\right)\nonumber\\
&+&\frac{1}{2}\int \Phi'(\Phi'')^2dx.  \label{eq51}
\end{eqnarray}
The last term cannot be integrated in exact form. Thus, I have shown that the equilibrium solution cannot be extended to $O(\epsilon^4)$.

\subsection{Stationary solution of the FPE including $O(\epsilon^3)$ terms}
Since I cannot find an equilibrium solution of the reduced equation for $R$ that holds beyond $O(\epsilon^2)$ terms, I go back to the full FPE \eqref{eq11} and find its stationary solution including terms beyond this order. Let me start with the nondimensional version of Eq.~\eqref{eq9} and try to find an approximation to the stationary solution of the FPE \eqref{eq11}.
\begin{eqnarray}
\rho_s&=&\frac{1}{Z}\exp\!\left[-\!\left(\frac{p^2}{2}(1+\epsilon^2\Phi'')+\Phi +\frac{\epsilon^2}{2}\Phi'^2\right)\!\right]\nonumber\\
&\times& [1+\epsilon^2r(x,p;\epsilon)].\label{eq52}
\end{eqnarray}
I use the extra term $\epsilon^2r(x,p;\epsilon)$ to cancel the term proportional to $\epsilon^3 p^3 \Phi'''$ that precludes finding an equilibrium solution to the full FPE, cf. section \ref{sec:2}. Inserting Eq.~\eqref{eq52} into Eq.~\eqref{eq11}, I obtain after some simplification,
\begin{eqnarray}
\epsilon^2\!\left(\frac{\partial^2r}{\partial p^2}-p\frac{\partial r}{\partial p}\right)\!+\epsilon^3p\Phi'''\!\left(\frac{p^2}{2}-1\right)\nonumber\\ -\epsilon^3\!\left(p\frac{\partial r}{\partial x}-\Phi'\frac{\partial r}{\partial p}\right)\! =O(\epsilon^4).   \label{eq53}
\end{eqnarray}
Assuming that $r=a(x)+\epsilon b(x)p+\epsilon c(x)p^3+O(\epsilon^2)$, Eq.~\eqref{eq53} yields
\begin{eqnarray}
&&\epsilon^3p(6c-b-a')+\epsilon^3p^3\!\left(\frac{\Phi'''}{2}-3c\right)\!= O(\epsilon^4)\nonumber\\
&&\Longrightarrow c=\frac{1}{6}\Phi''',\quad a'+b=\Phi'''.   \label{eq54}
\end{eqnarray}
A simple choice is $b=0$, which produces the stationary nonequilibrium probability density:
\begin{eqnarray}
\rho_s(x,p;\epsilon)&=&\frac{1}{Z}\exp\!\left[-\!\left(\frac{p^2}{2}(1+\epsilon^2\Phi'')+\Phi +\frac{\epsilon^2}{2}\Phi'^2\right)\!\right]\quad\nonumber\\
&\times&\!\left(1+\epsilon^2\Phi''+\frac{\epsilon^3}{6}p^3\Phi'''+ O(\epsilon^4)\right)\!.\label{eq55}
\end{eqnarray}
Selecting $a=3\Phi''/2$, $b=-\epsilon\Phi'''/2$, yields 
\begin{widetext}
\begin{eqnarray}
\rho_s(x,p;\epsilon)\!&=&\!\frac{\exp\!\left[-\!\left(\frac{p^2}{2}(1+\epsilon^2\Phi'')+\Phi +\frac{\epsilon^2}{2}\Phi'^2\right)\!\right]\!}{Z}\!\left(1+\frac{3\epsilon^2}{2}\Phi''\!+\frac{\epsilon^3(p^3-3p)}{6}\Phi'''\!+ O(\epsilon^4)\right)\quad\nonumber\\
\!&=&\!\frac{1}{Z}e^{-\frac{p^2}{2}-\Phi}\!\left(1-\frac{\epsilon^2}{2}[\Phi'^2+(p^2-3)\Phi'']+\frac{\epsilon^3(p^3-3p)}{6}\Phi'''\!+ O(\epsilon^4)\right)\!,   \label{eq56}
\end{eqnarray}
which, for one AOUP, is the approximate probability density in Eq.~(6) of Ref.~\cite{fodor}.
\end{widetext}

\section{Conclusions}\label{sec:4}
For zero persistence time, active Ornstein-Uhlenbeck particles become overdamped particles in contact with a thermal bath. Thus, they reach equilibrium for long times. There has been some controversy on whether AOUPs under harmonic potentials may reach a time invariant equilibrium state for nonzero persistence times \cite{fodor,mandal,cap_comment,pug17,dab19,sei19}. For nonzero persistence time $\tau$, AOUPs reach an equilibrium state characterized by a probability density that is invariant under time reversal if, and only if, their interaction potential is quadratic (within the class of smooth potentials). This can be shown by means of a general formula for their time reversed Langevin-Ito stochastic differential equation (the time reversal theorem, \cite{hau86}). Using path integrals, Fodor {\em et al} have concluded that AOUPs reach equilibrium with zero production of entropy \cite{fodor}, which agrees with the previous result. Mandal {\em et al} \cite{mandal} have disputed this conclusion using an incorrect time-reversed Langevin-Ito equation of the AOUPs that is at odds with the time reversal theorem. Their formulas for production of entropy and thermodynamics arguments are based upon their time reversed stochastic equation \cite{mandal}, and should be appropriately modified \cite{pug17}. 

The active harmonic oscillator under an additional thermal noise reaches a nonequilibrium stationary state that is no longer invariant under time reversal, cf. Ref.~\cite{dab19} and also Appendix \ref{ap5} in this paper. Thus, the model of the noisy overdamped particle in a harmonic potential is quite peculiar. It has an equilibrium probability density if the particle is subject to only one external noise, either Ornstein-Uhlenbeck or thermal white noise (the usual case for a purely passive particle), but it reaches a nonequilibrium stationary state when both noises are present. Other models for active colloidal particles include both translational thermal white noise and orientational white noise for the active velocity, which render the models thermodynamically consistent \cite{sei19}. 

For general interaction potentials and in the limit of small persistence time, AOUPs are close to an extended equilibrium state. In this paper, the state of affairs is made clear by using a Chapman-Enskog expansion in a dimensionless parameter $\epsilon\propto\sqrt{\tau}$. I have shown that the AOUP probability density is asymptotic to a local equilibrium in the momenta times a reduced probability density $R$ that depends on space and time. The continuity equation for the latter contains a probability current that depends on $\epsilon^2$. Its leading and first order terms depend only on first and second spatial derivatives of $R$. To $O(\epsilon^2)$, there is an equilibrium solution $R\propto e^{-\tilde{\Phi}}$ whose probability current vanishes. The $O(\epsilon^4)$ term in the probability current includes fifth order spatial derivatives of $R$ and there is no longer an equilibrium probability density [approximate to $O(\epsilon^4)$] that makes the current zero. The overall momentum-dependent stationary probability density has $O(\epsilon^3)$ terms that are odd in the momenta \cite{fodor}. Thus, this density is not invariant under time reversal. 

\acknowledgments
This work has been supported by the FEDER/Ministerio de Ciencia, Innovaci\'on y Universidades -- Agencia Estatal de Investigaci\'on grant MTM2017-84446-C2-2-R. I thank Jonathan Goodman  and John Neu for fruitful discussions, and Russel Caflisch for hospitality during a sabbatical stay at the Courant Institute. 

\appendix
\section{Time reversed diffusion process}\label{ap1}
Consider the system of Langevin-Ito equations
\begin{equation}
dX_t= b(X_t,t)\, dt+ \sigma(X_t,t)\, dw_t, \label{a1}
\end{equation}
where $X_t$ and $b(X_t,t)$ take on values in $\mathbb{R}^n$, $t\in(0,t_f)$, and $\sigma(X_t,t)$ is a $n\times l$ matrix with $l\leq n$. Under mild hypothesis, there is a time reversed diffusion process $\overline{X}_t=X_{t_f-t}$ that satisfies the equations \cite{hau86}
\begin{eqnarray}
&&d\overline{X}_t= \overline{b}(\overline{X}_t,t)\, dt+ \overline{\sigma}(\overline{X}_t,t)\, d\overline{w}_t, \label{a2}\\
&&\overline{b}^i(x,t)=-b^i(x,t_f-t)\nonumber\\
&&\quad +\frac{1}{\rho(x,t_f-t)}\frac{\partial}{\partial x_j}[a^{ij}(x,t_f-t)\rho(x,t_f-t)], \nonumber\\
&& \overline{a}^{ij}(x,t)=a^{ij}(x,t_f-t),\,\, a(x,t)=\sigma(x,t)\sigma(x,t)^T.\quad \label{a3}
\end{eqnarray}
Here superscripts denote components of vectors, summation over repeated indices is implied, and $\rho(x,t)$ is the solution of the FPE corresponding to Eq.~\eqref{a1}:
\begin{eqnarray}
\frac{\partial\rho}{\partial t}+\frac{\partial}{\partial x_i}\!\left[b^i(x,t)\rho-\frac{1}{2}\frac{\partial}{\partial x_j}(a^{ij}(x,t)\rho)\right]\!=0. \label{a4}
\end{eqnarray}
The proof shows that the infinitesimal generator of the time reversed process $\overline{X}_t$ is given by the coefficient functions in Eq.~\eqref{a3}. This is done straightforwardly by using two test functions and integration by parts. Then the resulting formulas are justified in the appropriate functional spaces under mild conditions for the coefficient functions \cite{hau86}. Stratonovich integration shows that the noise $\hat{w}_t=\overline{w}_{t_f-t}$ satisfies
\begin{eqnarray}
\hat{w}_t^i\!\!=\!w_t^i\!-\!w_{t_f}^i\!\!-\!\int_t^{t_f}\!\!\frac{ds}{\rho(X_s,s)}\frac{\partial}{\partial x_j}[\sigma^{ji}(X_s,s)\rho(X_s,s)],\quad \label{a5}
\end{eqnarray}
cf. Remark 2.5 in Ref.~\cite{hau86}. Note that selecting $d\overline{t}=d(t_f-t)>0$, $\overline{b}(x,t)$ changes sign in Eq.~\eqref{a2}, which is the same convention with positive differentials  $dt>0$ used in Refs.~\cite{nel67,che15} and elsewhere in the present paper.

\section{Systems of active Ornstein-Uhlenbeck particles}\label{ap2}
The AOUP equation of motion is \cite{fodor}
\begin{equation}
\dot{\mathbf{r}}_i=-\mu\nabla_i\Phi+\mathbf{v}_i,\quad \tau\dot{\mathbf{v}}_i=-\mathbf{v}_i+\sqrt{2D}\eta_i(t). \label{b1}
\end{equation}
where $i=1,\dots N$, and $\eta_i(t)$ is a zero-mean delta-correlated white noise. From this equation and $\mathbf{p}_i=\dot{\mathbf{r}}_i$, it follows
\begin{eqnarray*}
\dot{\mathbf{p}}_i&=&\dot{\mathbf{v}}_i-\mu\sum_{k=1}^N\dot{\mathbf{r}}_k\cdot\nabla_k\nabla_i\Phi\\
&=&-\frac{\mathbf{v}_i}{\tau}-\mu\sum_{k=1}^N(\dot{\mathbf{r}}_k\cdot\nabla_k)\nabla_i\Phi+\sqrt{\frac{2D}{\tau^2}}\eta_i(t).
\end{eqnarray*}
In terms of $\mathbf{p}_i$, the system of equations is
\begin{eqnarray}
&&\dot{\mathbf{r}}_i=\mathbf{p}_i,\label{b2}\\
&&\dot{\mathbf{p}}_i\!=\!-\frac{\mathbf{p}_i\!+\!\mu\nabla_i\Phi}{\tau}\!-\!\mu\sum_{k=1}^N\!(\mathbf{p}_k\!\cdot\nabla_k)\nabla_i\Phi\!+\!\sqrt{\frac{2D}{\tau^2}}\eta_i(t).\quad\label{b3}
\end{eqnarray}
The corresponding FPE for the probability density $\rho(\mathbf{R},\mathbf{P},t)$ (in which $\mathbf{R}=\mathbf{r}_1,\ldots, \mathbf{r}_N$, with a similar meaning for $\mathbf{P}$) is 
\begin{widetext}
\begin{equation}
\frac{\partial \rho}{\partial t}+\frac{\partial}{\partial r_{i\alpha}}(p_{i\alpha}\rho)-\frac{\partial}{\partial p_{i\alpha}}\left[\left(\frac{p_{i\alpha}}{\tau} +\frac{\mu}{\tau} (1+\tau p_{j\beta}\frac{\partial}{\partial r_{j\beta}})\frac{\partial}{\partial r_{i\alpha}}\Phi \right)\! \rho +\frac{D}{\tau^2}\frac{\partial \rho}{\partial p_{i\alpha}}\right]\!=0. \label{b4}
\end{equation}
Here $\alpha=1,2,\ldots, d$ are the components of the vectors $\mathbf{r}_i$ and $\mathbf{p}_i$. In Eq.~\eqref{b4}, summation over repeated indices is intended.
\end{widetext}

According to Eqs.~\eqref{a2}-\eqref{a4} in Appendix \ref{ap1}, the reverse time stochastic differential equations of Eqs.~\eqref{b2}-\eqref{b3} for $dt>0$ are \cite{nel67,hau86,che15}
\begin{eqnarray}
&&\dot{\overline{\mathbf{r}}}_i=\mathbf{\overline{p}}_i,\label{b5}\\
&&\dot{\mathbf{\overline{p}}}_i=-\frac{\mathbf{\overline{p}}_i+\mu\nabla_i\Phi(\overline{\mathbf{R}})}{\tau}-\mu\sum_{k=1}^N(\mathbf{\overline{p}}_k\cdot\nabla_k)\nabla_i\Phi(\overline{\mathbf{R}})\nonumber\\
&&\quad-\frac{2D}{\tau^2}\frac{\partial\ln\rho}{\partial\mathbf{p}_i}+\sqrt{\frac{2D}{\tau^2}}\overline{\eta}_i(t).\label{b6}
\end{eqnarray}
In equilibrium, $\rho=\rho_s(\mathbf{R},\mathbf{P})$ is invariant under time reversal. This  means that the drift term of Eq.~\eqref{b6} for $\rho=\rho_s$ is the same as the drift term of Eq.~\eqref{b3} under the time reversal transformation: $t\to-t$, $\mathbf{\overline{r}}_i=\mathbf{r}_i$, $\mathbf{\overline{p}}_i=-\mathbf{p}_i$:
\begin{eqnarray*}
-\frac{\mathbf{p}_i+\mu\nabla_i\Phi}{\tau}-\mu\sum_{k=1}^N(\mathbf{p}_k\cdot\nabla_k)\nabla_i\Phi-\frac{2D}{\tau^2}\frac{\partial\ln\rho_s}{\partial\mathbf{p}_i}\\
=\frac{\mathbf{p}_i-\mu\nabla_i\Phi}{\tau}+\mu\sum_{k=1}^N(\mathbf{p}_k\cdot\nabla_k)\nabla_i\Phi.
\end{eqnarray*}
This yields
\begin{eqnarray}
&&\frac{2D}{\tau^2}\frac{\partial\ln\rho_s}{\partial\mathbf{p}_i}=-\frac{2}{\tau}\!\left[\mathbf{p}_i+\mu\tau\sum_{k=1}^N(\mathbf{p}_k\cdot\nabla_k)\nabla_i\Phi\right]\Longrightarrow \nonumber\\
&&\rho_s\!=\!\exp\!\left[\!-\frac{\tau}{2D}\!\sum_{i=1}^N\![\mathbf{p}_i^2\!+\!\mu\tau(\mathbf{p}_i\cdot\nabla_i)^2\Phi]\!+\! \Lambda(\mathbf{r}_1,\ldots,\mathbf{r}_N)\!\right]\!.\quad\,\,\label{b7}
\end{eqnarray}
I now insert this stationary probability density in the FPE \eqref{b4} thereby finding
\begin{eqnarray*}
&&\sum_{i=1}^N\!\left[\mathbf{p}_i\cdot\nabla_i\Lambda+\frac{\mu}{D}\mathbf{p}_i\cdot\nabla_i\Phi+\frac{\mu^2\tau}{D}\sum_{j=1}^N (\nabla_j \Phi\cdot\nabla_j)(\mathbf{p}_i\cdot\nabla_i)\Phi\right]\quad\\
&&\times\rho_s-\frac{\mu\tau^2\rho_s}{2D}\sum_{i=1}^N(\mathbf{p}_i\cdot\nabla_i)^3\Phi =0, 
\end{eqnarray*}
where the first term can be cancelled by choosing $\Lambda=-\frac{\mu}{D}\Phi-\frac{\mu^2\tau}{2D}\sum_{j=1}^N|\nabla_j \Phi|^2$ but not the second. Thus, there is an equilibrium state
\begin{widetext}
\begin{eqnarray}
\rho_s=\frac{1}{Z}\exp\!\left[-\frac{\mu}{D}\!\left(\Phi +\frac{\mu\tau}{2}\sum_{j=1}^N|\nabla_j \Phi|^2\right)\!-\frac{\tau}{2D}\sum_{i=1}^N[\mathbf{p}_i^2+\mu\tau(\mathbf{p}_i\cdot\nabla_i)^2\Phi]\right]\!,\label{b8}
\end{eqnarray}\end{widetext}
if and only if $\sum_{i=1}^N(\mathbf{p}_i\cdot\nabla_i)^3\Phi =0$, which occurs for a smooth quadratic potential. I conclude that the stationary solution of the FPE (with smooth potential $\Phi$) is not an equilibrium invariant under time reversal unless $\Phi$ is quadratic. 

\section{Reduced probability density for a system of $N$ AOUPs}\label{ap3}
Here I nondimensionalize the model according to Table \ref{table2}. 
\begin{table}[ht]
\begin{center}\begin{tabular}{cccccc}
 \hline
$r_{i\alpha}$& $p_{i\alpha}$ & $t$ &$\Phi$& $P$\\ 
$l=\frac{L}{N^{1/d}}$ & $\sqrt{\frac{D}{\tau}}$ & $l\sqrt{\frac{\tau}{D}}$ &$\frac{D}{\mu}=T$ & $\sqrt{\frac{\tau}{Dl^2}}$\\ 
\hline
\end{tabular}
\end{center}
\caption{Units for nondimensionalizing the equations of the model. $L^d$ is box volume. }
\label{table2}
\end{table}
The nondimensional FPE corresponding to Eq.~\eqref{b4} is  
\begin{widetext}
\begin{equation}
\frac{\partial}{\partial p_{i\alpha}}\!\left[\left(p_{i\alpha}+\frac{\partial}{\partial p_{i\alpha}}\right)\rho\right]\!= \epsilon\left[\frac{\partial\rho}{\partial t}+\frac{\partial}{\partial r_{i\alpha}}(p_{i\alpha}\rho)-\frac{\partial\Phi}{\partial r_{i\alpha}}\frac{\partial\rho}{\partial p_{i\alpha}}-\epsilon \frac{\partial^2\Phi}{\partial r_{i\alpha}\partial r_{j\beta}}\,\frac{\partial }{\partial p_{i\alpha}}(p_{j\beta}\rho)\right]\!, \label{c1}
\end{equation}\end{widetext}
where summation over repeated indices is implied and
\begin{equation}
\epsilon=\frac{\sqrt{D\tau}}{l}=\sqrt{D\tau} \rho_n^{1/d}\ll 1, \quad\rho_n=\frac{N}{L^d}. \label{c2}
\end{equation}
Assuming the diffusive length is much smaller than the specific particle length $l=(L^d/N)^{1/d}$  corresponds to the limit $\sqrt{\tau}\to 0$ in Ref.~\cite{fodor}. Note that for the parameters listed in Fig.~1 of Ref.~\cite{fodor}, $\epsilon$ is large, 17.89 ($l=L/\sqrt{N}=2.5$) or 22.36 ($l=2$, the range of the repulsive potential), so that motility induced separation occurs in the opposite limit of large particle density. For $\epsilon=0$, $\mathbf{R}=(\mathbf{r}_1,\ldots,\mathbf{r}_N)$, $\mathbf{P}=(\mathbf{p}_1,\ldots,\mathbf{p}_N)$, Eq.~\eqref{c1} has the solution
\begin{eqnarray}
\rho^{(0)}(\mathbf{R},\mathbf{P},t)= \frac{e^{-p_{i\alpha}^2/2}}{(2\pi)^{dN/2}}\rho(\mathbf{R},t;\epsilon),\label{c3}\\ 
\int P^{(0)}(\mathbf{R},\mathbf{P},t)\, d\mathbf{P}= \rho(\mathbf{R},t;\epsilon). \nonumber
\end{eqnarray}
The Chapman-Enskog ansatz is \cite{BT10}
\begin{eqnarray}
&& \rho(\mathbf{R},\mathbf{P},t;\epsilon)=\frac{e^{-p_{i\alpha}^2/2}}{(2\pi)^{dN/2}}R(\mathbf{R},t;\epsilon)\nonumber\\
&&\quad+\sum_{j=1}^\infty\epsilon^j\rho^{(j)}(\mathbf{R},\mathbf{P};R), \label{c4}\\
&&\frac{\partial R}{\partial t}= \sum_{j=1}^\infty\epsilon^j\mathcal{F}^{(j)}(R),\label{c5}
\end{eqnarray}
where 
\begin{eqnarray}
&&\int \rho^{(j)}(\mathbf{R},\mathbf{P};R)\, d\mathbf{P}=0 \mbox{ for $j\geq 1$,} \label{c6}\\
&&\int R(\mathbf{R},t;\epsilon)\, d\mathbf{R}=1. \nonumber
\end{eqnarray}
Integrating Eq.~\eqref{c1} over the momenta, I obtain the exact continuity equation \begin{equation}
\frac{\partial R}{\partial t}\!+\! \frac{\partial J_{i\alpha}}{\partial r_{i\alpha}}\!=\! 0, \, J_{i\alpha}(\mathbf{R},t;\epsilon)\!=\!\int p_{i\alpha} \rho(\mathbf{R},\mathbf{P},t;\epsilon) d\mathbf{P}.\label{c7}
\end{equation}

Insertion of \eqref{c3}-\eqref{c5} into \eqref{c1} produces the hierarchy of equations
\begin{widetext}
\begin{eqnarray}
&&\mathcal{L}\rho^{(1)}\equiv \frac{\partial}{\partial p_{i\alpha}}\left(p_{i\alpha}+\frac{\partial}{\partial p_{i\alpha}}\right)\!\rho^{(1)}= \frac{e^{-p_{j\beta}^2/2}}{(2\pi)^{dN/2}}\, p_{i\alpha}\left(\frac{\partial\Phi}{\partial r_{i\alpha}}+\frac{\partial}{\partial r_{i\alpha}} \right)R,\label{c8} \\
&& \mathcal{L}\rho^{(2)}=\frac{e^{-p_{j\beta}^2/2}}{(2\pi)^{dN/2}}\left[\mathcal{F}^{(1)}+ \frac{\partial^2\Phi}{\partial r_{i\alpha}\partial r_{j\beta}}\,(p_{i\alpha}p_{j\beta}-\delta_{ij}\delta_{\alpha\beta})\rho\right]\!+p_{i\alpha}\frac{\partial\rho^{(1)}}{\partial r_{i\alpha}}-\frac{\partial\Phi}{\partial r_{i\alpha}}\frac{\partial\rho^{(1)}}{\partial p_{i\alpha}}, \label{c9}\\
&& \mathcal{L}\rho^{(3)}=p_{i\alpha}\frac{\partial\rho^{(2)}}{\partial r_{i\alpha}} -\frac{\partial\Phi}{\partial r_{i\alpha}}\frac{\partial \rho^{(2)}}{\partial p_{i\alpha}}-\frac{\partial^2\Phi}{\partial r_{i\alpha}\partial r_{j\beta}}\,\frac{\partial(p_{j\beta}\rho^{(1)})}{\partial p_{i\alpha}}+\frac{\delta \rho^{(1)}}{\delta R}\mathcal{F}^{(1)}, \label{c10}\\
&& \mathcal{L}\rho^{(4)}=\frac{e^{-p_{j\beta}^2/2}}{(2\pi)^{\frac{dN}{2}}}\mathcal{F}^{(3)} +p_{i\alpha}\frac{\partial\rho^{(3)}}{\partial r_{i\alpha}}-\frac{\partial\Phi}{\partial r_{i\alpha}}\frac{\partial \rho^{(3)}}{\partial p_{i\alpha}}-\frac{\partial^2\Phi}{\partial r_{i\alpha}\partial r_{j\beta}}\,\frac{\partial(p_{j\beta}\rho^{(2)})}{\partial p_{i\alpha}}+\frac{\delta\rho^{(2)}}{\delta R}\mathcal{F}^{(1)}, \label{c11}\\
&& \mathcal{L}\rho^{(5)}=p_{i\alpha}\frac{\partial\rho^{(4)}}{\partial r_{i\alpha}} -\frac{\partial\Phi}{\partial r_{i\alpha}}\frac{\partial\rho^{(4)}}{\partial p_{i\alpha}}-\frac{\partial^2\Phi}{\partial r_{i\alpha}\partial r_{j\beta}}\,\frac{\partial(p_{j\beta}\rho^{(3)})}{\partial p_{i\alpha}}+\frac{\delta \rho^{(1)}}{\delta R}\mathcal{F}^{(3)}+\frac{\delta\rho^{(3)}}{\delta R}\mathcal{F}^{(1)}, \label{c12}
\end{eqnarray}\end{widetext}
etc. The solvability conditions for the equations of this hierarchy are that the integrals over momenta of their right hand sides be zero. The solutions $P^{(j)}$ of the hierarchy are Gaussians in $p_{i\alpha}$ times polynomials of degree $j$: 
\begin{eqnarray}
&&\rho^{(2j+\xi)}(\mathbf{R},\mathbf{P};\rho)=\frac{e^{-p_{z\omega}^2/2}}{(2\pi)^{dN/2}}\nonumber\\
&&\times\sum_{n=0}^j\mathcal{A}^{(2j+\xi)}_{i_1\alpha_1,\ldots,i_{2n+\xi}\alpha_{2n+\xi}}\prod_{k=1}^{2n+\xi}p_{i_k\alpha_k}, \label{c13}
\end{eqnarray}
where $\xi=0,1$. Clearly for Eqs.~\eqref{c8}-\eqref{c12}, the solvability conditions yield
\begin{eqnarray}
&&\mathcal{F}^{(j)}=-\frac{\partial J^{(j)}_{i\alpha}}{\partial r_{i\alpha}},\quad J^{(j)}_{i\alpha} =\int p_{i\alpha}\rho^{(j)} d\mathbf{P}.\label{c14}\\
&&\mbox{Therefore, } J^{(2j)}_{i\alpha}=0,\quad \mathcal{F}^{(2j)}=0,\nonumber
\end{eqnarray} 
which I have used this to suppress all such terms in the previous hierarchy of equations. Eqs. \eqref{c14} agree with the continuity equation \eqref{c7}. Using 
\begin{equation}
\mathcal{L}\!\left(\frac{e^{-p_{j\beta}^2/2}}{(2\pi)^{dN/2}} p_{i\alpha}\right)\!=\!-\frac{e^{-p_{j\beta}^2/2}}{(2\pi)^{dN/2}} p_{i\alpha},\label{c15}
\end{equation}
The solution of Eq.~\eqref{c8} that satisfies Eq.~\eqref{c6} is:
\begin{widetext}
\begin{equation}
\rho^{(1)}=- \frac{e^{-p_{j\beta}^2/2}}{(2\pi)^{dN/2}} p_{i\alpha}\! \left(\frac{\partial\Phi}{\partial r_{i\alpha}}+\frac{\partial}{\partial r_{i\alpha}} \right)\!\rho\equiv- \frac{p_{i\alpha}e^{-p_{j\beta}^2/2}}{(2\pi)^{dN/2}}\,\mathcal{D}_{i\alpha}\rho. \label{c16}
\end{equation}
Inserting this in \eqref{c9}, its solvability condition produces
\begin{equation}
\mathcal{F}^{(1)}=-\frac{\partial J^{(1)}_{i\alpha}}{\partial r_{i\alpha}},\quad J^{(1)}_{i\alpha}=\int p_{i\alpha}\rho^{(1)}d\mathbf{P}=-\mathcal{D}_{i\alpha}R,\quad \mathcal{D}_{i\alpha}= \frac{\partial\Phi}{\partial r_{i\alpha}}+\frac{\partial}{\partial r_{i\alpha}}. \label{c17}
\end{equation}
Using
\begin{eqnarray}
\mathcal{L}\!\left(\frac{e^{-p_{j\beta}^2/2}}{(2\pi)^{dN/2}} p_{i\alpha}p_{j\beta}\right)\!=\!-\frac{e^{-p_{j\beta}^2/2}}{(2\pi)^{dN/2}} 2(p_{i\alpha}p_{j\beta}-\delta_{ij}\delta_{\alpha\beta}),\,\,\int \frac{e^{-p_{j\beta}^2/2}}{(2\pi)^{dN/2}}p_{i\alpha}p_{j\beta} d\mathbf{P}=\delta_{ij}\delta_{\alpha\beta},\quad \label{c18}
\end{eqnarray}
the solution of Eq.~\eqref{c9} that satisfies Eq.~\eqref{c6}:
\begin{eqnarray}
&&\rho^{(2)}=\frac{e^{-p_{j\beta}^2/2}}{(2\pi)^{dN/2}} (p_{i\alpha}p_{j\beta}-\delta_{ij}\delta_{\alpha\beta})\, \mathcal{A}^{(2)}_{i\alpha,j\beta},\label{c19}\\
&&\mathcal{A}^{(2)}_{i\alpha,j\beta}=\frac{1}{2}\!\left(\frac{\partial^2R}{\partial r_{i\alpha}\partial r_{j\beta}}+\frac{\partial\Phi}{\partial r_{i\alpha}}\frac{\partial R}{\partial r_{j\beta}}+\frac{\partial\Phi}{\partial r_{j\beta}}\frac{\partial R}{\partial r_{i\alpha}}+ \frac{\partial\Phi}{\partial r_{i\alpha}}\frac{\partial\Phi}{\partial r_{j\beta}}R\right)\!. \label{c20}
\end{eqnarray}
Eq.~\eqref{c19} has the form Eq.~\eqref{c13} with $\mathcal{A}^{(2)}=-\mathcal{A}^{(2)}_{i\alpha,i\alpha}$. 

Let me find $\rho^{(3)}$. Using 
\begin{equation}
\mathcal{L}\!\left(\frac{e^{-p_{j\beta}^2/2}}{(2\pi)^{\frac{dN}{2}}} p_{i\alpha}p_{j\beta}p_{k\gamma}\right)\!=\!\frac{e^{-p_{j\beta}^2/2}}{(2\pi)^{\frac{dN}{2}}}(2\delta_{ij}\delta_{\alpha\beta}p_{k\gamma} +2 \delta_{ik}\delta_{\alpha\gamma}p_{j\beta}+2 \delta_{jk}\delta_{\beta\gamma}p_{i\alpha} -3p_{i\alpha}p_{j\beta}p_{k\gamma}),\label{c21}
\end{equation}
I find the solution of Eq.~\eqref{c10} that satisfies Eq.~\eqref{c6}: 
\begin{eqnarray}
&&\rho^{(3)}\!\!=\!\frac{p_{i\alpha}e^{-p_{j\beta}^2/2}}{(2\pi)^{dN/2}}(\mathcal{A}^{(3)}_{(i\alpha,j\beta,k\gamma)} p_{j\beta}p_{k\gamma}\! +\! \mathcal{A}^{(3)}_{i\alpha})\! =\! \frac{p_{i\alpha}e^{-p_{j\beta}^2/2}}{(2\pi)^{dN/2}}[\mathcal{A}^{(3)}_{(i\alpha,j\beta,k\gamma)}\!(p_{j\beta}p_{k\gamma}\!-\!3\delta_{jk}\delta_{\beta\gamma}) \!+\!J^{(3)}_{i\alpha}],\,\,\quad \label{c22}\\
&&\mathcal{A}^{(3)}_{(i\alpha,j\beta,k\gamma)}= \frac{1}{3}(\mathcal{A}^{(3)}_{i\alpha,j\beta,k\gamma} +\mathcal{A}^{(3)}_{i\alpha,k\gamma,j\beta} +\mathcal{A}^{(3)}_{j\beta,k\gamma,i\alpha}),\label{c23}\\
&&\mathcal{A}^{(3)}_{i\alpha,j\beta,k\gamma} =-\frac{1}{3}\!\left(\!\mathcal{D}_{k\gamma}\mathcal{A}^{(2)}_{i\alpha,j\beta}-\frac{\partial^2\Phi}{\partial r_{i\alpha}\partial r_{j\beta}}\!\mathcal{D}_{k\gamma}R\right)\!,\label{c24}\\
&&\mathcal{A}^{(3)}_{i\alpha}\! =\! \!\left(\!\mathcal{D}_{i\alpha}\frac{\partial}{\partial r_{j\beta}}\!-\!\frac{\partial^2\Phi}{\partial r_{i\alpha}\partial r_{j\beta}\!}\right)\!\mathcal{D}_{j\beta}R\! -\frac{\partial^2\Phi}{\partial r_{j\beta}^2}\mathcal{D}_{i\alpha}R\!+\!6\mathcal{A}^{(3)}_{(i\alpha,j\beta,j\beta)}\! +\!2\frac{\partial\Phi}{\partial r_{j\beta}}\mathcal{A}^{(2)}_{i\alpha,j\beta}\! +\!\mathcal{D}_{i\alpha}\mathcal{A}^{(2)}_{j\beta,j\beta}.\,\,\quad\label{c25}
\end{eqnarray}
The corresponding contribution to the probability current is
\begin{eqnarray}
J^{(3)}_{i\alpha}\!&=&3\mathcal{A}^{(3)}_{(i\alpha,j\beta,j\beta)}+\mathcal{A}^{(3)}_{i\alpha}= \!\left(\mathcal{D}_{i\alpha}\frac{\partial}{\partial r_{j\beta}} +\frac{\partial^2\Phi}{\partial r_{i\alpha}\partial r_{j\beta}}\right)\!\mathcal{D}_{j\beta}R - 2\frac{\partial\mathcal{A}^{(2)}_{i\alpha,j\beta}}{\partial r_{j\beta}}\Longrightarrow\nonumber\\
 J^{(3)}_{i\alpha}\!&=& \frac{\partial}{\partial r_{j\beta}}\!\left(\frac{\partial^2\Phi}{\partial r_{i\alpha}\partial r_{j\beta}}R\right)\!. \label{c26}
\end{eqnarray}

The solution of Eq.~\eqref{c11} with conditions given Eq.~\eqref{c6} has the form of Eq.~\eqref{c13}:
\begin{eqnarray}
&&\rho^{(4)}\!=\!\frac{e^{-p_{j\beta}^2/2}}{(2\pi)^{dN/2}}\{\mathcal{A}^{(4)}_{(i\alpha,j\beta,k\gamma,l\delta)}(p_{i\alpha}p_{j\beta}p_{k\gamma}p_{l\delta}-\delta_{ij}\delta_{\alpha\beta}\delta_{kl}\delta_{\gamma\delta}-\delta_{ik}\delta_{\alpha\gamma}\delta_{jl}\delta_{\beta\delta}-\delta_{il}\delta_{\alpha\delta}\delta_{jk}\delta_{\beta\gamma})\nonumber\\
&&\quad\quad+\mathcal{A}^{(4)}_{i\alpha,j\beta}(p_{i\alpha}p_{j\beta}-\delta_{ij}\delta_{\alpha\beta})\},\label{c27}\\
&&\mathcal{A}^{(4)}_{i\alpha,j\beta,k\gamma,l\delta}=-\frac{1}{4}\!\left(\mathcal{D}_{i\alpha}\mathcal{A}^{(3)}_{(j\beta,k\gamma,l\delta)}+ \frac{\partial^2\Phi}{\partial r_{i\alpha}\partial r_{j\beta}}\mathcal{A}^{(2)}_{k\gamma,l\delta}\right)\!,\label{c28}\\
&&\mathcal{A}^{(4)}_{(i\alpha,j\beta)}\!=\!\frac{1}{4}\!\left(\frac{\partial^2}{\partial r_{i\alpha}\partial r_{j\beta}} +\frac{\partial\Phi}{\partial r_{j\beta}}\frac{\partial}{\partial r_{i\alpha}}+\frac{\partial\Phi}{\partial r_{i\alpha}}\frac{\partial}{\partial r_{j\beta}}\! +\frac{\partial\Phi}{\partial r_{i\alpha}}\frac{\partial\Phi}{\partial r_{j\beta}}\right)\!\frac{\partial}{\partial r_{k\gamma}}\mathcal{D}_{k\gamma}R\quad\quad \nonumber\\
&&\quad\quad +\!\frac{3}{2}\frac{\partial\Phi}{\partial r_{(k\gamma}}\mathcal{A}^{(3)}_{(i\alpha,j\beta,k\gamma))}+\frac{1}{2}\frac{\partial^2\Phi}{\partial r_{k\gamma}^2}\mathcal{A}^{(2)}_{i\alpha,j\beta}-\frac{1}{4}(\mathcal{D}_{i\alpha} J^{(3)}_{j\beta}+\mathcal{D}_{j\beta} J^{(3)}_{i\alpha}),\label{c29}\\
&&\mathcal{A}^{(4)}= -3\mathcal{A}^{(4)}_{(i\alpha,i\alpha,j\beta,j\beta)\!} -\mathcal{A}^{(4)}_{i\alpha,i\alpha},   \label{c30}
\end{eqnarray}
to derive which I have used 
\begin{eqnarray}
&&\mathcal{L}\!\left(\frac{e^{-p_{j\beta}^2/2}}{(2\pi)^{\frac{dN}{2}}} p_{i\alpha}p_{j\beta}p_{k\gamma}p_{l\delta}\right)\!=\!\frac{2e^{-p_{j\beta}^2/2}}{(2\pi)^{\frac{dN}{2}}}(\delta_{ij}\delta_{\alpha\beta}p_{k\gamma}p_{l\delta} +\delta_{ik}\delta_{\alpha\gamma}p_{j\beta}p_{l\delta}+\delta_{il}\delta_{\alpha\delta}p_{j\beta} p_{k\gamma}\nonumber\\
&&\quad\quad\quad\quad\quad\quad\quad\quad\quad\quad\quad+\delta_{jk}\delta_{\beta\gamma}p_{i\alpha} p_{l\delta}+\delta_{jl}\delta_{\beta\delta}p_{i\alpha} p_{k\gamma}+\delta_{kl}\delta_{\gamma\delta}p_{i\alpha}p_{j\beta}-2p_{i\alpha}p_{j\beta}p_{k\gamma}p_{l\delta}),\quad\label{c31}\\
&&\int \frac{e^{-p_{j\beta}^2/2}}{(2\pi)^{dN/2}}p_{i\alpha}p_{j\beta} p_{k\gamma}p_{l\delta}d\mathbf{P}=\delta_{ij}\delta_{\alpha\beta}\delta_{kl}\delta_{\gamma\delta}+\delta_{ik}\delta_{\alpha\gamma}\delta_{jl}\delta_{\beta\delta}+\delta_{il}\delta_{\alpha\delta}\delta_{jk}\delta_{\beta\gamma}.\label{c32}
\end{eqnarray}

After inserting Eqs.~\eqref{c16}, \eqref{c19}, \eqref{c22} and \eqref{c27} into Eq.~\eqref{c12}, the latter becomes
\begin{eqnarray}
&&\mathcal{L}\rho^{(5)}=\frac{e^{-p_{j\beta}^2/2}}{(2\pi)^{dN/2}}\!\left\{p_{i\alpha}\!\mathcal{D}_{i\alpha}[\mathcal{A}^{(4)}_{j\beta,k\gamma,l\delta,m\epsilon}(p_{j\beta}p_{k\gamma}p_{l\delta}p_{m\epsilon} -3\delta_{(jk}\delta_{\beta\gamma}\delta_{lm}\delta_{\delta\epsilon)})+\mathcal{A}^{(4)}_{j\beta,k\gamma}(p_{j\beta}p_{k\gamma}-\delta_{jk}\delta_{\beta\gamma})] \nonumber \right.\\
&&-2p_{i\alpha}\frac{\partial\Phi}{\partial r_{j\beta}}\!\left( 2\mathcal{A}^{(4)}_{(i\alpha,j\beta,k\gamma,l\delta)}p_{k\gamma}p_{l\delta}+ \mathcal{A}^{(4)}_{(i\alpha,j\beta)}\right)\! +\frac{\partial^2 \Phi}{\partial r_{i\alpha}\partial r_{j\beta}}\mathcal{A}^{(3)}_{k\gamma,l\delta,m\epsilon} [p_{i\alpha}p_{j\beta}p_{k\gamma}(p_{l\delta}p_{m\epsilon}- 3 \delta_{lm}\delta_{\delta\epsilon})]\nonumber\\
&&- \!\left(\frac{\partial^2 \Phi}{\partial r_{l\delta}^2}\mathcal{A}^{(3)}_{i\alpha,j\beta,k\gamma}+3 \frac{\partial^2 \Phi}{\partial r_{i\alpha}\partial r_{l\delta}}\mathcal{A}^{(3)}_{j\beta,k\gamma,l\delta}\right)\!p_{i\alpha}p_{j\beta}p_{k\gamma} +\frac{\partial^2 \Phi}{\partial r_{i\alpha}\partial r_{j\beta}}J^{(3)}_{k\gamma}(p_{i\alpha}p_{j\beta}p_{k\gamma} - \delta_{ij}\delta_{\alpha\beta}p_{k\gamma}\nonumber\\
&&-\delta_{ik}\delta_{\alpha\gamma}p_{j\beta}) + \frac{\delta\mathcal{A}^{(3)}_{(i\alpha,j\beta,k\gamma)}}{\delta\rho} \frac{\partial}{\partial r_{l\delta}}(\mathcal{D}_{l\delta}R) p_{i\alpha} (p_{j\beta}p_{k\gamma}-3\delta_{jk}\delta_{\beta\gamma}) + p_{i\alpha}\frac{\delta J^{(3)}_{i\alpha}}{\delta\rho} \frac{\partial}{\partial r_{l\delta}}\mathcal{D}_{l\delta}R \nonumber\\
&& -2\frac{\partial\Phi}{\partial r_{i\alpha}} (2\mathcal{A}^{(4)}_{(i\alpha,j\beta,k\gamma,l\delta)}p_{j\beta}p_{k\gamma}p_{l\delta}-\mathcal{A}^{(4)}_{(i\alpha,j\beta)}p_{j\beta})\left.+\mathcal{D}_{i\alpha}\frac{\partial J^{(3)}_{j\beta}}{\partial r_{j\beta}} p_{i\alpha}\!\right\}\!.\quad \label{c33}
\end{eqnarray}
Its solution is Eq.~\eqref{c13} with
\begin{eqnarray}
\mathcal{A}^{(5)}_{i\alpha,j\beta,k\gamma,l\delta,m\epsilon}\!&=&\!-\frac{1}{5}\!\left[\mathcal{D}_{i\alpha} \mathcal{A}^{(4)}_{j\beta,k\gamma,l\delta,m\epsilon}+ \frac{\partial^2 \Phi}{\partial r_{i\alpha}\partial r_{j\beta}} \mathcal{A}^{(3)}_{k\gamma,l\delta,m\epsilon}\right]\!,\label{c34}\\
\mathcal{A}^{(5)}_{i\alpha,j\beta,k\gamma}\!&=&\! \frac{1}{3}\!\left[20\mathcal{A}^{(5)}_{(i\alpha,j\beta,k\gamma,l\delta,l\delta)}\! -\mathcal{D}_{i\alpha} \mathcal{A}^{(4)}_{j\beta,k\gamma}\!+4\frac{\partial\Phi}{\partial r_{l\delta}}\mathcal{A}^{(4)}_{(i\alpha,j\beta,k\gamma,l\delta)}\!\right.\nonumber\\
\!&+&\left.\!\frac{\partial^2 \Phi}{\partial r_{l\delta}^2}\mathcal{A}^{(3)}_{i\alpha,j\beta,k\gamma}+ 3\frac{\partial^2\Phi}{\partial r_{i\alpha}\partial r_{l\delta}}\mathcal{A}^{(3)}_{(j\beta,k\gamma,l\delta)}\!-\frac{\partial^2\Phi}{\partial r_{i\alpha}\partial r_{j\beta}}\mathcal{A}^{(3)}_{k\gamma}\!-\frac{\delta\mathcal{A}^{(3)}_{i\alpha,j\beta,k\gamma}}{\delta\rho} \frac{\partial}{\partial r_{l\delta}}\mathcal{D}_{l\delta} R\right]\!\nonumber\\
\!&=&\! -\frac{1}{3}\!\left[\mathcal{D}_{i\alpha} \mathcal{A}^{(4)}_{j\beta,k\gamma}\!+4\frac{\partial}{\partial r_{(l\delta}}\mathcal{A}^{(4)}_{i\alpha,j\beta,k\gamma,l\delta)}\!+\!\frac{\partial^2\Phi}{\partial r_{i\alpha}\partial r_{j\beta}}\mathcal{A}^{(3)}_{k\gamma}\!+\frac{\delta\mathcal{A}^{(3)}_{i\alpha,j\beta,k\gamma}}{\delta\rho} \frac{\partial}{\partial r_{l\delta}}\mathcal{D}_{l\delta} R\right]\!,   \label{c35}\\
\mathcal{A}^{(5)}_{i\alpha}\!&=&\!6\mathcal{A}^{(5)}_{(i\alpha,j\beta,j\beta)} -\mathcal{D}_{i\alpha} \mathcal{A}^{(4)}+2\frac{\partial\Phi}{\partial r_{j\beta}} \mathcal{A}^{(4)}_{(i\alpha,j\beta)} +\frac{\partial^2 \Phi}{\partial r_{i\alpha} \partial r_{j\beta}}\mathcal{A}^{(3)}_{j\beta} \nonumber\\
\!&+&\!\frac{\partial^2 \Phi}{\partial r_{j\beta}^2}\mathcal{A}^{(3)}_{i\alpha}-\frac{\delta\mathcal{A}^{(3)}_{i\alpha}}{\delta\rho} \frac{\partial}{\partial r_{j\beta}}\mathcal{D}_{j\beta}R +\mathcal{D}_{i\alpha} \mathcal{F}^{(3)}.\quad \label{c36}
\end{eqnarray}
To find these coefficients, I have used the following formulas:
\begin{eqnarray}
&&\mathcal{L}\!\left(\frac{e^{-p_{j\beta}^2/2}}{(2\pi)^{\frac{dN}{2}}} p_{i\alpha}p_{j\beta}p_{k\gamma}p_{l\delta}p_{m\epsilon}\right)\!=\!\frac{e^{-p_{j\beta}^2/2}}{(2\pi)^{\frac{dN}{2}}}(2\delta_{ij}\delta_{\alpha\beta}p_{k\gamma}p_{l\delta}p_{m\epsilon} +2\delta_{ik}\delta_{\alpha\gamma}p_{j\beta}p_{l\delta}p_{m\epsilon}+2\delta_{il}\delta_{\alpha\delta}p_{j\beta} p_{k\gamma}p_{m\epsilon}\nonumber\\
&&\quad\quad\quad\quad\quad\quad\quad\quad\quad\quad\quad\quad+2\delta_{im}\delta_{\alpha\epsilon}p_{j\beta}p_{k\gamma}p_{l\delta}+2\delta_{jk}\delta_{\beta\gamma}p_{i\alpha} p_{l\delta}p_{m\epsilon}+2\delta_{jl}\delta_{\beta\delta}p_{i\alpha} p_{k\gamma}p_{m\epsilon}\nonumber\\
&&\quad\quad\quad\quad\quad\quad\quad\quad\quad\quad\quad\quad
+2\delta_{jm}\delta_{\beta\epsilon}p_{i\alpha} p_{k\gamma}p_{l\delta}+2\delta_{kl}\delta_{\gamma\delta}p_{i\alpha}p_{j\beta}p_{m\epsilon}+2\delta_{km}\delta_{\gamma\epsilon}p_{i\alpha} p_{j\beta}p_{l\delta}\nonumber\\
&&\quad\quad\quad\quad\quad\quad\quad\quad\quad\quad\quad\quad +2\delta_{lm}\delta_{\delta\epsilon}p_{i\alpha} p_{j\beta} p_{k\gamma}-5p_{i\alpha}p_{j\beta}p_{k\gamma}p_{l\delta}p_{m\epsilon}),\label{c37}
\end{eqnarray}
and 
\begin{eqnarray}
&&\int \frac{e^{-p_{j\beta}^2/2}}{(2\pi)^{dN/2}}p_{i\alpha}p_{j\beta} p_{k\gamma}p_{l\delta}p_{m\epsilon}p_{n\phi}d\mathbf{P}=\delta_{ij}\delta_{\alpha\beta}\delta_{kl}\delta_{\gamma\delta}\delta_{mn}\delta_{\epsilon\phi}+\delta_{ik}\delta_{\alpha\gamma}\delta_{jl}\delta_{\beta\delta}\delta_{mn}\delta_{\epsilon\phi}\nonumber\\
&&+\delta_{il}\delta_{\alpha\delta}\delta_{jk}\delta_{\beta\gamma}\delta_{mn}\delta_{\epsilon\phi}+ \delta_{im}\delta_{\alpha\epsilon}\delta_{jk}\delta_{\beta\gamma}\delta_{ln}\delta_{\delta\phi}+ \delta_{in}\delta_{\alpha\phi}\delta_{jk}\delta_{\beta\gamma}\delta_{lm}\delta_{\delta\epsilon}+ \delta_{im}\delta_{\alpha\epsilon}\delta_{jl}\delta_{\beta\delta}\delta_{kn}\delta_{\gamma\phi}\nonumber\\
&&+ \delta_{in}\delta_{\alpha\phi}\delta_{jl}\delta_{\beta\delta}\delta_{km}\delta_{\gamma\epsilon}+\delta_{il}\delta_{\alpha\delta}\delta_{jm}\delta_{\beta\epsilon}\delta_{kn}\delta_{\gamma\phi}+\delta_{il}\delta_{\alpha\delta}\delta_{jn}\delta_{\beta\phi}\delta_{km}\delta_{\gamma\epsilon}+\delta_{ik}\delta_{\alpha\gamma}\delta_{jm}\delta_{\beta\epsilon}\delta_{ln}\delta_{\delta\phi}\nonumber\\
&&+ \delta_{ij}\delta_{\alpha\beta}\delta_{km}\delta_{\gamma\epsilon}\delta_{ln}\delta_{\delta\phi}+\delta_{ij}\delta_{\alpha\beta}\delta_{kn}\delta_{\gamma\phi}\delta_{ln}\delta_{\epsilon\delta}+\delta_{im}\delta_{\alpha\epsilon}\delta_{jn}\delta_{\beta\phi}\delta_{kl}\delta_{\gamma\delta}+\delta_{ik}\delta_{\alpha\gamma}\delta_{jn}\delta_{\beta\phi}\delta_{lm}\delta_{\delta\epsilon}\nonumber\\
&&+\delta_{in}\delta_{\alpha\phi}\delta_{jm}\delta_{\beta\phi}\delta_{kl}\delta_{\gamma\delta}.\label{c38}
\end{eqnarray}

The probability current that corresponds to $\rho^{(5)}$ is
\begin{eqnarray}
&&J^{(5)}_{i\alpha} =72 \mathcal{A}^{(5)}_{(i\alpha,j\beta,j\beta,k\gamma,k\gamma)}\! +2\frac{\partial\Phi}{\partial r_{k\gamma}} (6\mathcal{A}^{(4)}_{(i\alpha,j\beta,j\beta,k\gamma)} + \mathcal{A}^{(4)}_{(i\alpha,k\gamma)})- \mathcal{D}_{i\alpha} (2\mathcal{A}^{(4)}_{j\beta,j\beta}-3\mathcal{A}^{(4)}_{(j\beta,j\beta,k\gamma,k\gamma)})\!\nonumber\\ 
&&\quad\quad  +\, 6\, \frac{\partial^2\Phi}{\partial r_{i\alpha}\partial r_{j\beta}} \mathcal{A}^{(3)}_{(k\gamma,k\gamma,j\beta)}+ \frac{\partial^2\Phi}{\partial r_{j\beta}^2} J^{(3)}_{i\alpha}+\frac{\partial^2\Phi}{\partial r_{i\alpha}\partial r_{j\beta}} J^{(3)}_{j\beta}-\mathcal{D}_{i\alpha} \frac{\partial J^{(3)}_{j\beta}}{\partial r_{j\beta}}\!- \frac{\delta J^{(3)}_{i\alpha}}{\delta R}\frac{\partial}{\partial r_{j\beta}}\mathcal{D}_{j\beta} R.\quad \label{c39}
\end{eqnarray}

The resulting reduced equation for the probability density $\rho$ is
\begin{eqnarray}
\frac{\partial R}{\partial(\epsilon t)}=\frac{\partial}{\partial r_{i\alpha}}\!\left[R\frac{\partial\Phi}{\partial r_{i\alpha}}+\frac{\partial R}{\partial r_{i\alpha}} -\epsilon^2\frac{\partial}{\partial r_{j\beta}}\!\left(R\frac{\partial^2\Phi}{\partial r_{i\alpha}\partial r_{j\beta}} \right)\!-\epsilon^4 J^{(5)}_{i\alpha}\right]\!+ O(\epsilon^6)\equiv -\frac{\partial J^r_{i\alpha}}{\partial r_{i\alpha}}. \label{c40}
\end{eqnarray}
Note that the reduced probability density evolves in a slow time scale $\epsilon t$.
\end{widetext}

\section{Equilibrium solution}\label{ap4}
There is an equilibrium solution that solves the reduced equation with $J^r_{i\alpha}=0$ to order $O(\epsilon^2)$, with $J^r_{i\alpha}$ given by Eq.~\eqref{c40}. I find it by inserting the exponential form
\begin{equation}
\rho_{\rm eq}=e^{\tilde{f}_{\rm eq}-\tilde{\Phi}}, \quad\tilde{\Phi}(\mathbf{R};\epsilon)=\Phi(\mathbf{R})+\sum_{j=1}^\infty \epsilon^{2j}\Phi^{(2j)}(\mathbf{R}),\label{d1}
\end{equation}
into $J^r_{i\alpha}=0$, and then finding $\Phi^{(2j)}$. The reduced free energy $\tilde{f}_{\rm eq}$ ensures that the normalization condition \eqref{c6} is fulfilled. To leading order, $J^{(1)}_{i\alpha}=0$ produces $\tilde{\Phi}=\Phi$ according to Eq.~\eqref{c17}. Keeping the $\epsilon^2$ term in $J^r_{i\alpha}$ given by Eq.~\eqref{c40}, I get
\begin{eqnarray*}
-\epsilon^2\!\left(\frac{\partial\Phi^{(2)}}{\partial r_{i\alpha}}+\frac{\partial^3\Phi}{\partial r_{i\alpha}\partial r_{j\beta}^2}-\frac{\partial\Phi}{\partial r_{j\beta}}\frac{\partial^2\Phi}{\partial r_{i\alpha}\partial r_{j\beta}}\right)\!=O(\epsilon^4),
\end{eqnarray*}
which yields
\begin{eqnarray}
\rho_{\rm eq}\!=\!e^{\tilde{f}_{\rm eq}-\tilde{\Phi}}, \,\tilde{\Phi}\!=\!\Phi\! +\!\epsilon^2\!\!\left[\frac{1}{2}\!\left(\frac{\partial\Phi}{\partial r_{i\alpha}}\right)^2\!\!-\!\frac{\partial^2\Phi}{\partial r_{i\alpha}^2}\right]\!+\!O(\epsilon^4).\quad\quad  \label{d2}
\end{eqnarray}

\section{AOUP in a thermal bath}\label{ap5}
What happens in the AOUPs are placed in a thermal bath? Dabelow {\em et al} claim that thermodynamic considerations can only be made for AOUPs in a thermal bath because the behavior of the active velocity $v$ under time reversal is ambiguous \cite{dab19}. To elucidate this point, I consider the simple case of an active harmonic oscillator in a thermal bath and show that it evolves to a nonequilibrium stationary state. Curiously, one can write this stationary state as a Gaussian that satisfies the equipartition theorem for  appropriately defined momentum and effective temperature, cf. Eq.~\eqref{e6} below. The SDEs are 
\begin{eqnarray}
\dot{x}=-\mu\kappa x+v+\sqrt{2 D_x}\,\eta_x(t),\, \tau\dot{v}=-v+\sqrt{2D}\,\eta(t),\quad\quad \label{e1}
\end{eqnarray}
where $D_x=\mu T$ is the diffusivity due to the bath at temperature $T$ in units of energy. Firstly, consider the nondimensional version of Eq.~\eqref{e1} when units are defined as in Table \ref{table3}.
\begin{table}[ht]
\begin{center}\begin{tabular}{cccc}
 \hline
$x$& $v$ & $t$ \\ 
$\sqrt{D\tau}$ & $\sqrt{\frac{D}{\tau}}$ & $\tau$ \\ 
\hline
\end{tabular}
\end{center}
\caption{Units for nondimensionalizing Eq.~\eqref{e1}. }
\label{table3}
\end{table}
Eq.~\eqref{e1} becomes
\begin{eqnarray}
\dot{x}=-k x+v+\sqrt{2d}\,\eta_x(t),\quad \dot{v}=-v+\sqrt{2}\,\eta(t), \label{e2}
\end{eqnarray}
where $k=\mu\kappa\tau$, $d=D_x/D$,  and the $\eta$s are independent zero mean Gaussian white noises with correlations $\delta(t-t')$. The stationary probability density for Eq.~\eqref{e2} is Gaussian, proportional to $\exp(-\underline{x}^T\underline{\underline{\mathcal{M}}}^{-1}\underline{x}/2)$, in which $\underline{\underline{\mathcal{M}}}$ is the matrix of the second order correlations and $\underline{x}=(x,v)$. The correlation matrix is calculated by imposing that $\underline{\underline{\mathcal{L}}}\,\underline{\underline{\mathcal{M}}}+\underline{\underline{\mathcal{M}}}\,\underline{\underline{\mathcal{L}}}^T =- 2\underline{\underline{\mathcal{D}}}$, where $\underline{\underline{\mathcal{L}}}$ is the coefficient matrix in Eq.~\eqref{e2} and the diffusion matrix $\underline{\underline{\mathcal{D}}}$ is diagonal with elements $d$ and 1 \cite{gha17}. The resulting correlations are 
\begin{eqnarray}
\langle x^2\rangle=\frac{d}{k}+\frac{1}{k(1+k)}, \quad\langle xv\rangle=\frac{1}{1+k},\quad\langle v^2\rangle =1.\quad \label{e3}
\end{eqnarray}
Straightforward computation produces the Gaussian density
\begin{eqnarray}
&&\rho_s=\frac{1}{Z}e^{-\beta\mathcal{H}},\label{e4}\\
&&\beta\mathcal{H}=\frac{1}{2}(\underline{x}^T\underline{\underline{\mathcal{M}}}^{-1}\underline{x})\nonumber\\
&&= \frac{1+k}{1+d(1+k)^2\!}\!\left(\frac{1+d+kd}{2}v^2-kxv+\frac{1+k}{2}kx^2\right)\!.\,\, \quad\label{e5}
\end{eqnarray}
After some more simple algebra, I find the following normal form of the quadratic energy:
\begin{eqnarray}
\mathcal{H}\!=\!\frac{1}{2}\frac{1\!+\!d\!+\!kd}{1\!+\!d(1+k)^2}\!\left(v\!-\!\frac{kx}{1\!+\!d\!+\!kd}\right)^2\!+\frac{1}{2}\frac{k x^2}{1\!+\!d\!+\!kd},\quad \label{e6}
\end{eqnarray}
with $\beta=1+k.$ In dimensional units, Eq.~\eqref{e6} becomes 
\begin{eqnarray}
&&\mathcal{H}=\frac{1+(1\!+\!\mu\tau\kappa)\frac{D_x}{D}}{1\!+\!(1\!+\!\mu\tau\kappa)^2\frac{D_x}{D}}\,\frac{\tau p^2}{2\mu}+\frac{1}{2}\frac{\kappa x^2}{1\!+\!(1\!+\!\mu\tau\kappa)\frac{D_x}{D}}, \label{e7}\\ 
&&p=v-\frac{\mu\kappa x}{1+(1+\mu\tau\kappa)\frac{D_x}{D}},\quad T_\text{eff}=\frac{D}{\mu(1+\mu\tau\kappa)}. \label{e8}
\end{eqnarray}
For $D_x=0$, Eqs.~\eqref{e7}-\eqref{e8} yield the formulas in Eq.~\eqref{eq10}.

Eqs.~\eqref{e4}-\eqref{e6} have the form of an equilibrium stationary density for an effective temperature $1/\beta$, and a Hamiltonian that is even in the momentum $p=v-kx/(1+d+kd)$. However, the AOUP subject to external white noise is out of equilibrium. Trying to apply the time reversal argument used in Section \ref{sec:2} provides some intuition about this. In terms of the variables $x$ and $p=v-kx$, Eqs.~\eqref{e2} are 
\begin{eqnarray}
&&\dot{x}=p+\sqrt{2d}\,\eta_x(t),\nonumber\\ 
&&\dot{p}=-kx-(1+k)p+\sqrt{2(1+k^2d)}\,\eta_p(t), \label{e9}
\end{eqnarray}
and their reversed stochastic differential equations are
\begin{eqnarray}
&&\dot{\overline{x}}= \overline{p}-2d\,\frac{\partial\ln\rho}{\partial x}(\overline{x},\overline{p},t)+\sqrt{2d}\,\overline{\eta}_x(t),\nonumber\\ 
&&\dot{\overline{p}}=-k\overline{x} -(1+k)\overline{p}-2(1+k^2d)\,\frac{\partial\ln\rho}{\partial p}(\overline{x},\overline{p},t)\nonumber\\ 
&&\quad+\sqrt{2(1+k^2d)}\,\overline{\eta}_p(t). \label{e10}
\end{eqnarray}
Here I have used that the variance of the combined white noise $\sqrt{2}\,\eta(t)-k\sqrt{2d}\,\eta_x(t)$ is $2(1+k^2d)$. If $d=0$, the first equation \eqref{e9} becomes the first equation \eqref{e10} under the time reversal transformation: $t\to-t$, $\overline{x}=x$, $\overline{p}=-p$. Then I can compare the drift terms of direct and time reversed second equations for an equilibrium probability density, which is also invariant under the time reversal transformation. If $d\neq 0$, {\em there is no longer a time reversal transformation that converts the first equations in Eqs.~\eqref{e9} and \eqref{e10} into each other}. Contrary to the case $d=0$, time reversal does not restrict the form of the (possibly nonequilibrium) stationary probability density.

To prove that the AOUP in a thermal bath is indeed out of equilibrium, I first show that the  stochastic area swept by the particle in the $x$-$v$ plane of Eq.~\eqref{e2} increases linearly with time \cite{gha17}. In fact, the time average of the stochastic area is 
\begin{eqnarray}
A&=&\lim_{t\to\infty}\frac{1}{2t}\int_0^t (v\dot{x}-x\dot{v})\, dt= (\mathcal{L}\mathcal{M}+\mathcal{D})_{12}\nonumber\\
&=&\langle xv\rangle=(1+k)^{-1}\neq 0.\label{e11}
\end{eqnarray}
A system with detailed balance having an invertible diffusion matrix satisfies $\underline{\underline{\mathcal{L}}}\,\underline{\underline{\mathcal{M}}}+\underline{\underline{\mathcal{D}}}=\underline{\underline{0}}$ \cite{gha17}. Thus Eq.~\eqref{e4} is a {\em nonequilibrium} stationary probability density. This argument does not apply to systems with non invertible diffusion matrix, which is the case for the harmonic AOUP with $d=0$.


\begin{thebibliography}{99}
\bibitem{fodor}E. Fodor, C. Nardini, M. E. Cates, J. Tailleur, P. Visco, and F. van Wijland, How Far from Equilibrium Is Active Matter? Phys. Rev. Lett. {\bf 117}, 038103 (2016).
\bibitem{gas04} P. Gaspard, Time-Reversed Dynamical Entropy and Irreversibility in Markovian Random Processes. J. Stat. Phys. {\bf 117}, 599-615 (2004).
\bibitem{and07} D. Andrieux, P. Gaspard, S. Ciliberto, N. Garnier, S. Joubaud, and A. Petrosyan, Entropy Production and Time Asymmetry in Nonequilibrium Fluctuations. Phys. Rev. Lett. {\bf 98}, 150601 (2007).
\bibitem{spi12} R. Spinney and I. Ford, Fluctuation Relations: A Pedagogical Overview, in {\em Nonequilibrium Statistical Physics of Small Systems: Fluctuation Relations and Beyond}, edited by R. Klages, W. Just, and C. Jarzynski. Wiley-VCH, Weinheim, 2012. 
\bibitem{mandal} D. Mandal, K. Klymko, and M. R. DeWeese, Entropy Production and Fluctuation Theorems for Active Matter. Phys. Rev. Lett. {\bf 119}, 258001 (2017).
\bibitem{cap_comment}L. Caprini, U. M. B. Marconi, A. Puglisi, and A. Vulpiani, Comment on Entropy Production and Fluctuation Theorems for Active Matter. Phys. Rev. Lett. {\bf 121}, 139801 (2018).
\bibitem{pug17} A. Puglisi and U. M. B. Marconi, Clausius Relation for Active Particles: What Can We Learn from Fluctuations. Entropy {\bf 19}(7), 356 (2017). 
\bibitem{dab19}L. Dabelow, S. Bo, and R. Eichhorn, Irreversibility in active matter systems: Fluctuation theorem and mutual information. Phys. Rev. X {\bf 9}, 021009 (2019).
\bibitem{hau86} U. G. Haussmann and E. Pardoux, Time reversal of diffusions. Ann. Probab. {\bf 14}, 1188-1205 (1986).
\bibitem{nel67} E. Nelson, Dynamical theories of Brownian motion, 2nd ed. Princeton U.P. Princeton, N.J., 1967.
\bibitem{che15} Y. Chen, T. Georgiou, and M. Pavon, Fast cooling for a system of stochastic oscillators. J. Math. Phys. {\bf 56}, 113302 (2015).
\bibitem{BT10} L. L. Bonilla and S. W. Teitsworth, Nonlinear wave methods for charge transport. Wiley-VCH, Weinheim, 2010.
\bibitem{gardiner} C. W. Gardiner, Stochastic methods. A handbook for the natural and social sciences. 4th ed. Springer, Berlin 2010.
\bibitem{cha70} S. Chapman and T. G. Cowling, The mathematical theory of non-uniform gases. 3rd ed. Cambridge U. P., Cambridge 1970.
\bibitem{sei19} U. Seifert, From Stochastic Thermodynamics to Thermodynamic Inference. Annu. Rev. Condens. Matter Phys. {\bf 10}, 171-192 (2019).
\bibitem{bon00} L. L. Bonilla, Chapman-Enskog method and synchronization of globally coupled oscillators. Phys. Rev. E {\bf 62}, 4862-4868 (2000).
\bibitem{BT19} L. L. Bonilla and C. Trenado, Contrarian compulsions produce exotic time-dependent flocking of active particles. Phys. Rev. E {\bf 99}, 012612 (2019). 
\bibitem{gha17}A. Ghanta, J. C. Neu, and S. W. Teitsworth, Fluctuation loops in noise-driven linear dynamical systems. Phys. Rev. E {\bf 95}, 032128 (2017).
\end{thebibliography}
\end{document}